\documentclass[usenatbib]{mn2e}
\usepackage{natbib}
\usepackage{amssymb}
\usepackage{epsfig}
\usepackage{rotating}
\usepackage{aas_macros}
\usepackage{longtable}
\usepackage{url}

\newcommand{\Lx}{\ensuremath{L_{\mathrm{X}}}}
\newcommand{\Lxt}{\ensuremath{L_{\mathrm{X,tot}}}}
\newcommand{\Lxc}{\ensuremath{L_{\mathrm{X,c}}}}
\newcommand{\kT}{\ensuremath{kT}}
\newcommand{\kTt}{\ensuremath{kT_{\mathrm{tot}}}}
\newcommand{\kTc}{\ensuremath{kT_{\mathrm{c}}}}
\newcommand{\Yx}{\ensuremath{Y_{\mathrm{X}}}}

\newcommand{\cshift}{\ensuremath{\langle {\mathrm w} \rangle}}

\newcommand{\Msol}{\ensuremath{\mathrm{M_{\odot}}}}

\newcommand{\rt}{\ensuremath{R_{\mathrm{200}}}}
\newcommand{\rf}{\ensuremath{R_{\mathrm{500}}}}

\newcommand{\Mf}{\ensuremath{M_{\mathrm{500}}}}

\newcommand{\OM}{\ensuremath{\Omega_{\mathrm{M}}}}

\newcommand{\Fcore}{\ensuremath{F_{\mathrm{core}}}}
\newcommand{\sigT}{\ensuremath{\sigma_{\mathrm{T}}}}

\newcommand{\etal}{et al.\ }

\newcommand{\ie}{{\it i.e.\ }}

\newcommand{\ASCA}{\emph{ASCA}}
\newcommand{\Chandra}{\emph{Chandra}}

\newcommand{\ROSAT}{\emph{ROSAT}}
\newcommand{\XMM}{\emph{XMM-Newton}}

\newcommand{\chisq}{\ensuremath{\chi^2}}

\newcommand{\gta}{\,\rlap{\raise 0.4ex\hbox{$>$}}{\lower 0.6ex\hbox{$\sim$}}\,}
\newcommand{\lta}{\,\rlap{\raise 0.4ex\hbox{$<$}}{\lower 0.6ex\hbox{$\sim$}}\,}

\newcommand{\nm}{\mbox{\ensuremath{\mathrm{~\nm}}}}

\newcommand{\cm}{\mbox{\ensuremath{\mathrm{~cm}}}}

\newcommand{\km}{\mbox{\ensuremath{\mathrm{~km}}}}

\newcommand{\kpc}{\mbox{\ensuremath{\mathrm{~kpc}}}}
\newcommand{\Mpc}{\mbox{\ensuremath{\mathrm{~Mpc}}}}
\newcommand{\s}{\mbox{\ensuremath{\mathrm{~s}}}}

\newcommand{\Gyr}{\mbox{\ensuremath{\mathrm{~Gyr}}}}

\newcommand{\keV}{\mbox{\ensuremath{\mathrm{~keV}}}}
\newcommand{\erg}{\mbox{\ensuremath{\mathrm{~erg}}}}

\newcommand{\cc}{\ensuremath{\mathrm{\cm^3}}}

\newcommand{\pcmsq}{\mbox{\ensuremath{\mathrm{~cm^{-2}}}}}
\newcommand{\pMpc}{\ensuremath{\mathrm{\Mpc^{-1}}}}
\newcommand{\ps}{\ensuremath{\mathrm{\s^{-1}}}}

\newcommand{\ergps}{\ensuremath{\mathrm{\erg \ps}}}
\newcommand{\flux}{\ensuremath{\mathrm{\erg \ps \pcmsq}}}

\newcommand{\kmpspMpc}{\ensuremath{\mathrm{\km \ps \pMpc\,}}}
\newcommand{\LT}{\mbox{\ensuremath{\mathrm{L_{X}-kT}}}}

\newcommand{\YM}{\mbox{\ensuremath{\mathrm{Y_{X}-M_{500}}}}}
\newcommand{\LM}{\mbox{\ensuremath{\mathrm{L_X-M}}}}

\newcommand{\MT}{\mbox{\ensuremath{\mathrm{M-kT}}}}

\newcommand{\LCDM}{$\Lambda$CDM~}

\begin{document}

\title[The self-similar \LT\ relation]{Self-similar scaling and evolution in the galaxy cluster
  X-ray Luminosity-Temperature relation.}
\author[B.J. Maughan \etal]
  {B. J. Maughan,$^{1}$\thanks{E-mail: ben.maughan@bristol.ac.uk}
    P. A. Giles,$^1$ S. W. Randall,$^2$ C. Jones,$^2$ and W. R. Forman$^2$\\
  $^1$H. H. Wills Physics Laboratory, University of Bristol, Tyndall Ave, Bristol BS8 1TL, UK.\\
  $^2$Harvard-Smithsonian Center for Astrophysics, 60 Garden St, Cambridge, MA 02140, USA.\\
}

\maketitle

\begin{abstract}
  We investigate the form and evolution of the X-ray
  luminosity-temperature (\LT) relation of a sample of 114 galaxy
  clusters observed with \Chandra\ at $0.1<z<1.3$. The clusters were
  divided into subsamples based on their X-ray morphology or whether
  they host strong cool cores. We find that when the core regions are
  excluded, the most relaxed clusters (or those with the strongest
  cool cores) follow an \LT\ relation with a slope that agrees well
  with simple self-similar expectations. This is supported by an
  analysis of the gas density profiles of the systems, which shows
  self-similar behaviour of the gas profiles of the relaxed clusters
  outside the core regions. By comparing our data with clusters in
  the REXCESS sample, which extends to lower masses, we find evidence
  that the self-similar behaviour of even the most relaxed clusters
  breaks at around $3.5\keV$. By contrast, the \LT\ slopes of the
  subsamples of unrelaxed systems (or those without strong cool cores)
  are significantly steeper than the self-similar model, with lower
  mass systems appearing less luminous and higher mass systems
  appearing more luminous than the self-similar relation. We argue
  that these results are consistent with a model of non-gravitational
  energy input in clusters that combines central heating with entropy
  enhancements from merger shocks. Such enhancements could extend the
  impact of central energy input to larger radii in unrelaxed
  clusters, as suggested by our data. We also examine the evolution of
  the \LT\ relation, and find that while the data appear inconsistent
  with simple self-similar evolution, the differences can be plausibly
  explained by selection bias, and thus we find no reason to rule out
  self-similar evolution. We show that the fraction of cool core
  clusters in our (non-representative) sample decreases at $z>0.5$ and
  discuss the effect of this on measurements of the evolution in the
  \LT\ relation.
\end{abstract}


\section{Introduction} \label{s.intro}
The study of galaxy clusters is often motivated by one of two
goals. First, clusters have been shown to be excellent cosmological
probes, providing complementary and competitive constraints on cosmological parameters
to those obtained using other techniques
\citep[e.g.][]{all08,vik09a}. Second, galaxy clusters are unique
laboratories in which to study extreme physical processes, such as
cluster mergers or the interaction between relativitic jets from
active galactic nuclei (AGN) and the X-ray emitting plasma of the
intra-cluster medium (ICM). In fact these goals are not
exclusive. Accurate cluster mass measurements are required for
cosmological studies, but these must be estimated from observable
properties which can be strongly affected by the non-gravitational
processes at play in clusters \citep{ran02,row04,kra06a,mag07,har08}.

The scaling relations between different cluster observables and
between those observables and cluster masses are a case in point. The
relationships between readily observable cluster properties such as
X-ray luminosity (\Lx) or temperature (kT) and mass can often be
approximated as simple power laws, and can be used to provide
``cheap'' mass estimates for large numbers of clusters out to high
redshifts \citep{mau07b}. Power law scaling relations are expected
under simplified models in which clusters are self-similar objects,
having formed in single monolithic gravitational collapses and whose
ICM is heated only by the shocks associated with the collapse
\citep{kai86}. In this self-similar model, galaxy clusters and groups
of all masses are identical objects when scaled by their mass. This is
referred to as {\em strong} self-similarity \citep{bow97}, and sets
the power law slopes of the scaling relations which are not
predicted to evolve with redshift. Evolution is expected, however, in
the normalisation of the scaling relations, and is due (in the
self-similar model) solely to the changing density of the Universe
with redshift \citep{bry98}. This redshift-dependent evolution in the
normalisation of the scaling relations is referred to as {\em weak}
self-similarity.

The \LT\ relation is the most well studied of these scaling relations
\citep[e.g.][]{mit79,mus84,edg91,mar98a,ett04,pra09a,mit11}, as these
two properties can easily be measured directly and essentially
independently from X-ray data. Indeed, the observational consensus is
that the slope of the \LT\ relation is $\sim3$, significantly steeper
than the self-similar prediction of $2$, and appears to steepen
further at galaxy group scales \citep[e.g.][]{hel00,osm04}. The slope
of the \LT\ relation is the archetypal example of the departure of
clusters from the self-similar predictions. Other evidence for
similarity breaking is seen in the radial distribution of the
ICM. Using simple models, \citet{neu99} found evidence for
self-similarity of cluster surface brightness profiles outside the
core regions, and \citet{arn02a} extended this result to higher
redshift clusters. \citet{pon99} showed that surface brightness
profiles of the hotter ($>4\keV$) clusters in their sample of relaxed
systems were self similar outside of the core regions. Recently, with
higher-quality data and more sophisticated modelling, the picture has
become more detailed; \citet{cro08} used \XMM\ observations to examine
the gas density profiles of a representative sample of 31 local
clusters and found that outside the core regions there is a
temperature dependence of the gas density profiles, but at larger
radii, the profiles become self-similar for the most massive
clusters. Using the same sample of clusters, \citep{arn10} also showed
that the pressure profiles of the ICM have a universal, self-similar
form, with low dispersion outside the core regions.

Perhaps the most detailed information on the processes responsible for
similarity breaking has come from observations of the entropy
distribution of the ICM. Entropy is useful this regard because it is
conserved in adiabatic processes, and so provides a good indicator of
the thermal history of the gas. A key observation has been that the
entropy profiles of galaxy groups and clusters show departures from a
self-similar power law, with profiles flattening in the core regions,
indicating an absence of the expected low-entropy gas in cluster cores
\citep[e.g.][]{pon99,voi05c,pra10}. As this low-entropy gas is also
the most luminous part of the ICM, a mechanism that removed or raised
the entropy of this gas, and that had a proportionately larger effect
in low mass systems, would also in principal explain the steeper than
self-similar slope of the \LT\ relation, and the similarity breaking in
ICM density profiles.

A great deal of work has been performed in investigating the different
processes that could be responsible for this similarity breaking, with
cooling out of the low entropy gas \citep[e.g.][]{bry00}, preheating
\citep[energy input at early times in the formation of clusters;
e.g.][]{bor02}, and energy input from supernovae-driven galactic
outflows all considered. A consensus is now emerging that energy input
from AGN at high redshift, when they are in quasar-mode with high
accretion rates, are responsible for providing a roughly constant
(with halo mass) level of entropy injection to the ICM, leading to the
observed ICM structure and scaling relations
\citep[e.g.][]{bow08,sho09,mcc11}.

Any study of the \LT\ relation must include a consideration of the
phenomenon of cool cores in clusters. These dense, cool regions in the
centres of many clusters radiate efficiently in X-rays, and were long
thought to be the sites of deposition of large quantities (hundreds of
solar masses per year) of condensed ICM \citep[see][for a
review]{fab94b}. High resolution X-ray spectroscopy subsequently
revealed that gas was not cooling out of the X-ray emitting phase in
the large quantities predicted, with the bulk of it instead
stabilising at around $1/3$ of the global temperature
\citep{pet01,kaa01,dav01}. This effectively substituted the problem of
explaining the fate of the cooling gas with the need for a mechanism
to balance the cooling process to give the observed, lower cooling
rates. This problem has been vigorously investigated over the past
decade, with energy input from AGN emerging as the most likely
candidate \citep[e.g.][]{mcn07}. This is a milder form of AGN input
than the quasar-mode heating that drives similarity breaking. Here the
energy is thought to be input to the ICM via the bubbles inflated by
the AGN jets and associated weak shocks, with the energy input coupled
to the cooling rate by some feedback mechanism to maintain a rough
balance between heating and cooling. Cool cores also have an important
observational effect in that they significantly increase (decrease)
the measured global \Lx\ (kT) compared to clusters without cool
cores. The presence and strength of cool cores can thus influence the
detectability of clusters in X-ray surveys, and complicate
measurements of the form and evolution of the \LT\ relation (and are a
dominant source of scatter in luminosity scaling relations).

Knowledge of the evolution of the \LT\ relation is important as a step
in the process of providing mass estimates for high-z clusters,
understanding the selection functions of X-ray cluster surveys, and
also for probing the history of the heating mechanisms in
clusters. For example, simulations have shown that AGN heating and
preheating models give divergent predictions for the evolution of the
\LT\ relation, allowing the models to be distinguished observationally
\citep{sho10}. Observational results are somewhat mixed, with some
studies finding that evolution to $z\sim1$ is consistent with
self-similar predictions in which the evolution is driven by the
increasing density of the Universe with redshift, leading to denser
collapsed objects \citep[e.g.][]{vik02,lum04c,mau06a}, while
other studies have found evidence for departures from self-similar
evolution \citep[e.g.][]{ett04,bra07}. Recent work, however, has
demonstrated the importance of considering selection bias in studies of
the evolution of \Lx\ scaling relations, showing that such biases can
mimic or reduce sensitivity to departures from self-similar evolution
\citep{sta06,pac07,nor08,man10a}.

In this paper we investigate the \LT\ relation of a sample of
114 clusters covering wide temperature ($2<\kT<16\keV$) and redshift
($0.1<z<1.3$) baselines to examine the strong and weak self similarity
of the cluster population. The sample was first presented in
\citet[][hereafter M08]{mau08a}, wherein the analysis methods were
described, and the evolution of the structural properties and metal
abundance were investigated. Subsequently the sample was used to show
that \Lx\ is a more precise mass estimator than had previously been
thought \citep{mau07b}. The current paper is organised as
follows: in \textsection 2 we review the sample and describe updates
to the analysis and calibration of the data since \citet{mau08a}; in
\textsection 3 we discuss the classification of clusters based on
dynamical and cool core state; in \textsection 4 and \textsection 5
respectively, strong and weak self-similarity are investigated, and in
\textsection 6 and \textsection 7 the results are discussed and our
conclusions are summarised. A \LCDM cosmology of $H_0=70\kmpspMpc
(\equiv100h\kmpspMpc$, and $\OM=0.3$ ($\Omega_\Lambda=0.7$) is adopted
throughout and all errors are quoted at the $68\%$ confidence level.

\section{Sample and data reduction}
The sample used for this study is a set of 114 galaxy clusters
covering the redshift range $0.1<z<1.3$ observed with \Chandra\
ACIS-I, and originally presented in M08.  In this work, the data have
been reanalysed using updated versions of the {\em CIAO} software
package (version 4.2), and the \Chandra\ Calibration Database (version
4.3.0). The full analysis procedure is as described in M08, but the
key steps are summarised in the following, along with any differences
in procedure from that work.

The data were reprocessed from the level 1 events and cleaned and
filtered. Blank-sky background files appropriate for each observation
were also prepared and normalised to match the $9.5-12\keV$ count rate
in the cluster data. Point sources were detected and excluded from all
further analysis. The X-ray centroid of the cluster was determined,
and radial profiles of the cluster and background data were used to
determine the extent of the cluster emission, and to define local
background regions free from cluster emission. Spectra were extracted
from the cluster and blank-sky files in these local background regions
and used to measure the difference in soft Galactic emission between
the cluster and blank-sky data. This difference spectrum was fit with
a soft APEC \citep{smi01} plasma model and included as a fixed
component in the spectral fits to the cluster spectra, with the
normalisation scaled by extraction area \citep[see][for
details]{vik05a}. Fits to the cluster spectra were performed in the
$0.6-9\keV$ band with an absorbed APEC model, with the absorbing
column fixed at the Galactic value \citep{dic90}.

The gas density profile of each cluster was determined by converting
the observed surface brightness profile (measured in the $0.7-2\keV$
band) into a projected emissivity profile, which was then modelled by
projecting a density model along the line of sight (see M08 for
details). The model used was that of \citet[][see that work for
definitions of the parameters]{vik06a};
\begin{eqnarray}\label{e.emis}
n_pn_e & = & \frac{n_0^2 (r/r_c)^{-\alpha}}{(1+r^2/r_c^2)^{3\beta-\alpha/2}} \times (1+r^\gamma/r_s^\gamma)^{-\epsilon/\gamma},
\end{eqnarray}
a modification of the widely used $\beta$-model, with added
flexibility to fit a power law cusp in the core and a change in slope
at large radii. Gas masses were then determined from Monte Carlo
realisations of the projected emissivity profile based on the best
fitting projected model to the original data. At each data point, a
new randomised point was drawn from a Gaussian distribution centered
on the model value at that point, with a standard deviation equal to
the fractional measurement error on the original data point,
multiplied by the model value. Note that this represents a minor change from
M08 in which we randomised the original {\em data} to derive the
uncertainties on the model parameters. The new approach is considered
superior as we avoid adding noise to an already noisy observed
profile.

The cluster temperature, gas mass and \rf\ (the radius enclosing a
mean density of 500 times the critical density at the cluster's
redshift) were then determined iteratively. The procedure followed was
to extract a spectrum from within an estimated \rf\ (with the central
$15\%$ of that radius excluded), integrate the gas density profile to
determine the gas mass within the estimated \rf, and thus calculate
\Yx \citep[the product of kT and the gas mass, and a low scatter proxy
for total mass][]{kra06a}. A new value of \rf\ was then estimated from
the \Yx-M scaling relation of \citet{vik09},
\begin{eqnarray}\label{e.ym}
\Mf & = & E^{-2/5}(z)A_{YM}\frac{Y_X}{3\times10^{14}\Msol\keV}^{B_{YM}},
\end{eqnarray}
with $A_{YM}=5.77\times10^{14}h^{1/2}\Msol$ and $B_{YM}=0.57$. Here,
\Mf\ is the mass within \rf\ (allowing \rf\ to be trivially computed),
and $E(z)=\sqrt{\OM(1+z)^3 + (1-\OM-\Omega_\Lambda)(1+z)^2 +
  \Omega_\Lambda}$, describing the redshift evolution of the Hubble
parameter. This \YM\ relation assumes self-similar evolution (as
$E^{-2/5}$), which has been shown to be a good description of observed
clusters to $z\approx0.6$ \citep{mau07b}.  Equation (\ref{e.ym}) is an
updated version of the \Yx-M relation used in M08, but the change is
negligible here, leading to a $\lta2\%$ increase in \rf\ for the range
of masses considered here. The process was repeated until \rf\
converged. The temperature and luminosity were then measured from
spectra extracted within \rf\ both with and without the central $15\%$
of that radius excluded. We use the notation \Lxt\ and \kTt\ to
indicate properties measured in the $(0-1)\rf$ aperture and \Lxc\ and
\kTc\ for those in the core-excised  $(0.15-1)\rf$ aperture.

The measured properties of the clusters are given in Table
\ref{t.props}. All luminosities are bolometric.

\subsection{Comparison with M08}
In addition to the general methodological differences between M08 and
the current work that are described above, some improvements were made
to the analyses of specific clusters. Specifically, for several
clusters additional light-

\clearpage
\onecolumn
\small
\begin{longtable}{lcccccc}
  \caption{Summary of cluster properties. \Lxt\ and \kTt\ were
    measured in the $(0-1)\rf$ aperture and \Lxc\ and \kTc\ were
    measured in the $(0.15-1)\rf$ aperture. Clusters are in order of
    right ascension, and luminosities are bolometric.}\label{t.props}\\
  \hline

Cluster & z & \rf\ & \kTt & \Lxt & \kTc & \Lxc \\
 & & (Mpc) & (keV) & ($10^{44}\ergps$) & (keV) & ($10^{44}\ergps$) \\ \hline
\endfirsthead

\caption{\emph{continued}}\\
\hline
Cluster & z & \rf\ & \kTt & \Lxt & \kTc & \Lxc \\
 & & (Mpc) & (keV) & ($10^{44}\ergps$) & (keV) & ($10^{44}\ergps$) \\ \hline
\endhead
\hline
\multicolumn{7}{r}{\emph{continued on next page}}
\endfoot
\hline
\endlastfoot

MS0015.9+1609	& 0.541	& 1.26	& $8.3^{+0.4}_{-0.3}$	& $50.6\pm0.6$	& $8.3^{+0.5}_{-0.4}$	& $35.5\pm0.5$ \\
RXJ0027.6+2616	& 0.367	& 0.99	& $5.2^{+1.3}_{-0.7}$	& $7.4\pm0.5$	& $4.8^{+1.0}_{-0.8}$	& $5.6\pm0.5$ \\
CLJ0030+2618	& 0.500	& 0.84	& $4.1^{+0.6}_{-0.8}$	& $4.9\pm0.9$	& $4.1^{+1.7}_{-1.0}$	& $3.6\pm0.8$ \\
A68	& 0.255	& 1.25	& $8.6^{+1.0}_{-0.5}$	& $17.7\pm0.5$	& $7.8^{+1.0}_{-1.0}$	& $10.4\pm0.4$ \\
A115	& 0.197	& 1.28	& $5.3^{+0.1}_{-0.1}$	& $13.7\pm0.1$	& $6.7^{+0.3}_{-0.3}$	& $9.7\pm0.1$ \\
A209	& 0.206	& 1.34	& $7.2^{+0.4}_{-0.4}$	& $19.2\pm0.3$	& $7.4^{+0.5}_{-0.5}$	& $13.1\pm0.2$ \\
CLJ0152.7-1357S	& 0.831	& 0.76	& $4.6^{+0.8}_{-0.7}$	& $9.1\pm0.6$	& $4.9^{+1.1}_{-0.9}$	& $7.1\pm0.6$ \\
A267	& 0.230	& 1.07	& $4.9^{+0.3}_{-0.3}$	& $12.4\pm0.6$	& $4.4^{+0.5}_{-0.4}$	& $7.3\pm0.5$ \\
CLJ0152.7-1357N	& 0.831	& 0.78	& $5.1^{+0.7}_{-0.7}$	& $11.8\pm0.6$	& $4.9^{+0.9}_{-0.8}$	& $9.6\pm0.8$ \\
MACSJ0159.8-0849	& 0.405	& 1.32	& $7.9^{+0.3}_{-0.3}$	& $42.0\pm0.5$	& $10.2^{+0.9}_{-0.9}$	& $19.0\pm0.4$ \\
CLJ0216-1747	& 0.578	& 0.78	& $5.9^{+2.9}_{-1.7}$	& $2.8\pm0.3$	& $5.6^{+3.8}_{-1.8}$	& $2.1\pm0.2$ \\
RXJ0232.2-4420	& 0.284	& 1.33	& $6.5^{+0.5}_{-0.4}$	& $30.7\pm0.9$	& $8.0^{+1.4}_{-1.1}$	& $16.7\pm0.7$ \\
MACSJ0242.5-2132	& 0.314	& 1.08	& $4.6^{+0.2}_{-0.2}$	& $28.4\pm0.7$	& $5.5^{+0.7}_{-0.6}$	& $7.8\pm0.3$ \\
A383	& 0.187	& 1.03	& $3.9^{+0.1}_{-0.1}$	& $9.9\pm0.2$	& $4.5^{+0.3}_{-0.3}$	& $3.8\pm0.1$ \\
MACSJ0257.6-2209	& 0.322	& 1.17	& $7.2^{+0.6}_{-0.5}$	& $17.3\pm0.4$	& $6.7^{+0.9}_{-0.6}$	& $8.7\pm0.3$ \\
MS0302.7+1658	& 0.424	& 0.81	& $3.4^{+0.5}_{-0.4}$	& $6.3\pm0.7$	& $3.3^{+0.8}_{-0.6}$	& $3.2\pm0.7$ \\
CLJ0318-0302	& 0.370	& 0.94	& $5.4^{+1.2}_{-0.9}$	& $6.0\pm0.4$	& $5.4^{+1.7}_{-1.2}$	& $3.9\pm0.3$ \\
MACSJ0329.6-0211	& 0.450	& 1.01	& $4.5^{+0.2}_{-0.3}$	& $28.6\pm0.8$	& $4.5^{+0.5}_{-0.4}$	& $11.6\pm0.5$ \\
MACSJ0404.6+1109	& 0.355	& 1.09	& $5.8^{+0.6}_{-0.5}$	& $10.8\pm0.4$	& $5.5^{+0.7}_{-0.5}$	& $8.7\pm0.4$ \\
MACSJ0429.6-0253	& 0.399	& 1.08	& $5.2^{+0.2}_{-0.2}$	& $24.5\pm0.6$	& $6.5^{+0.9}_{-0.7}$	& $9.3\pm0.4$ \\
RXJ0439.0+0715	& 0.230	& 1.14	& $5.4^{+0.3}_{-0.2}$	& $16.1\pm0.3$	& $5.3^{+0.4}_{-0.3}$	& $8.7\pm0.3$ \\
RXJ0439+0520	& 0.208	& 0.97	& $3.7^{+0.2}_{-0.2}$	& $9.3\pm0.4$	& $3.9^{+0.4}_{-0.4}$	& $3.3\pm0.3$ \\
MACSJ0451.9+0006	& 0.430	& 1.02	& $5.8^{+0.7}_{-0.9}$	& $16.9\pm0.8$	& $5.0^{+1.1}_{-0.6}$	& $11.1\pm0.7$ \\
A521	& 0.253	& 1.20	& $4.9^{+0.2}_{-0.2}$	& $16.4\pm0.3$	& $4.8^{+0.2}_{-0.2}$	& $13.8\pm0.3$ \\
A520	& 0.199	& 1.31	& $6.6^{+0.2}_{-0.2}$	& $18.4\pm0.2$	& $6.5^{+0.3}_{-0.3}$	& $14.5\pm0.2$ \\
MS0451.6-0305	& 0.550	& 1.19	& $6.8^{+0.8}_{-0.6}$	& $46.8\pm1.9$	& $7.6^{+1.2}_{-1.0}$	& $31.3\pm1.6$ \\
CLJ0522-3625	& 0.472	& 0.82	& $4.3^{+1.0}_{-0.9}$	& $3.3\pm0.3$	& $4.3^{+1.4}_{-1.0}$	& $2.5\pm0.2$ \\
CLJ0542.8-4100	& 0.642	& 0.91	& $6.4^{+0.8}_{-0.7}$	& $11.3\pm0.5$	& $6.2^{+1.0}_{-0.8}$	& $8.6\pm0.4$ \\
MACSJ0647.7+7015	& 0.584	& 1.26	& $10.9^{+1.4}_{-0.9}$	& $43.8\pm1.1$	& $11.3^{+2.1}_{-1.6}$	& $24.7\pm0.8$ \\
1E0657-56	& 0.296	& 1.64	& $11.5^{+0.4}_{-0.4}$	& $79.3\pm0.6$	& $11.7^{+0.5}_{-0.5}$	& $55.4\pm0.5$ \\
MACSJ0717.5+3745	& 0.546	& 1.40	& $11.2^{+0.7}_{-0.7}$	& $79.3\pm1.2$	& $10.6^{+1.0}_{-0.6}$	& $59.0\pm0.9$ \\
A586	& 0.171	& 1.23	& $7.2^{+0.5}_{-0.5}$	& $15.2\pm0.4$	& $7.6^{+0.8}_{-0.8}$	& $7.6\pm0.4$ \\
MACSJ0744.9+3927	& 0.697	& 1.08	& $7.7^{+0.4}_{-0.4}$	& $48.9\pm0.9$	& $8.1^{+0.7}_{-0.6}$	& $25.1\pm0.7$ \\
A665	& 0.182	& 1.42	& $7.5^{+0.3}_{-0.3}$	& $23.4\pm0.3$	& $7.8^{+0.4}_{-0.4}$	& $16.6\pm0.2$ \\
A697	& 0.282	& 1.49	& $9.8^{+0.5}_{-0.5}$	& $40.4\pm0.6$	& $10.2^{+0.8}_{-0.7}$	& $26.4\pm0.5$ \\
CLJ0848.7+4456	& 0.574	& 0.57	& $2.4^{+0.4}_{-0.3}$	& $1.1\pm0.2$	& $2.0^{+0.2}_{-0.3}$	& $0.9\pm0.2$ \\
ZWCLJ1953	& 0.320	& 1.14	& $7.0^{+0.6}_{-0.6}$	& $16.4\pm0.4$	& $6.1^{+0.6}_{-0.6}$	& $9.3\pm0.3$ \\
CLJ0853+5759	& 0.475	& 0.83	& $5.0^{+1.3}_{-0.9}$	& $3.0\pm0.4$	& $5.1^{+1.5}_{-0.9}$	& $2.6\pm0.4$ \\
MS0906.5+1110	& 0.180	& 1.10	& $5.2^{+0.2}_{-0.2}$	& $9.0\pm0.2$	& $4.7^{+0.3}_{-0.3}$	& $5.0\pm0.1$ \\
RXJ0910+5422	& 1.110	& 0.48	& $4.6^{+1.4}_{-1.1}$	& $2.8\pm0.3$	& $2.7^{+1.9}_{-0.8}$	& $2.2\pm0.4$ \\
A773	& 0.217	& 1.30	& $7.5^{+0.3}_{-0.3}$	& $18.3\pm0.3$	& $7.4^{+0.4}_{-0.4}$	& $11.4\pm0.2$ \\
A781	& 0.298	& 1.14	& $5.5^{+0.6}_{-0.6}$	& $11.8\pm0.5$	& $5.5^{+0.7}_{-0.5}$	& $9.9\pm0.5$ \\
CLJ0926+1242	& 0.489	& 0.84	& $4.0^{+0.5}_{-0.4}$	& $4.6\pm0.3$	& $4.5^{+1.0}_{-0.9}$	& $3.2\pm0.3$ \\
RBS797	& 0.354	& 1.19	& $6.2^{+0.3}_{-0.3}$	& $50.3\pm0.9$	& $7.3^{+1.0}_{-0.9}$	& $13.0\pm0.6$ \\
MACSJ0949.8+1708	& 0.384	& 1.24	& $7.5^{+0.6}_{-0.6}$	& $28.6\pm0.7$	& $7.3^{+0.9}_{-0.8}$	& $17.4\pm0.6$ \\
CLJ0956+4107	& 0.587	& 0.81	& $4.8^{+0.8}_{-0.7}$	& $5.8\pm0.3$	& $4.0^{+0.7}_{-0.5}$	& $4.5\pm0.4$ \\
A907	& 0.153	& 1.14	& $5.2^{+0.1}_{-0.1}$	& $11.5\pm0.1$	& $5.4^{+0.2}_{-0.2}$	& $5.1\pm0.1$ \\
MS1006.0+1202	& 0.221	& 1.12	& $5.6^{+0.4}_{-0.5}$	& $7.7\pm0.2$	& $6.0^{+0.5}_{-0.5}$	& $5.1\pm0.2$ \\
MS1008.1-1224	& 0.301	& 1.04	& $5.0^{+0.3}_{-0.3}$	& $10.4\pm0.3$	& $4.6^{+0.4}_{-0.4}$	& $6.7\pm0.3$ \\
ZW3146	& 0.291	& 1.28	& $6.2^{+0.1}_{-0.1}$	& $47.0\pm0.4$	& $7.8^{+0.4}_{-0.4}$	& $16.3\pm0.3$ \\
CLJ1113.1-2615	& 0.725	& 0.65	& $3.8^{+0.8}_{-0.6}$	& $3.8\pm0.4$	& $3.6^{+1.3}_{-0.8}$	& $2.4\pm0.5$ \\
A1204	& 0.171	& 0.96	& $3.4^{+0.1}_{-0.1}$	& $9.9\pm0.2$	& $3.7^{+0.3}_{-0.3}$	& $2.7\pm0.1$ \\
CLJ1117+1745	& 0.548	& 0.71	& $3.1^{+0.6}_{-0.5}$	& $2.1\pm0.3$	& $3.3^{+0.9}_{-0.6}$	& $1.8\pm0.2$ \\
CLJ1120+4318	& 0.600	& 0.93	& $5.8^{+0.9}_{-0.7}$	& $13.4\pm0.8$	& $5.2^{+1.3}_{-0.8}$	& $8.7\pm0.7$ \\
RXJ1121+2327	& 0.562	& 0.78	& $3.5^{+0.3}_{-0.3}$	& $5.0\pm0.3$	& $3.2^{+0.3}_{-0.3}$	& $4.3\pm0.3$ \\
A1240	& 0.159	& 0.92	& $3.8^{+0.3}_{-0.3}$	& $1.8\pm0.1$	& $3.8^{+0.3}_{-0.3}$	& $1.6\pm0.1$ \\
MACSJ1131.8-1955	& 0.307	& 1.43	& $8.1^{+0.9}_{-0.7}$	& $31.5\pm1.0$	& $9.5^{+1.8}_{-1.4}$	& $21.9\pm0.9$ \\
MS1137.5+6625	& 0.782	& 0.83	& $6.6^{+0.8}_{-0.7}$	& $14.0\pm0.5$	& $6.5^{+1.2}_{-1.0}$	& $7.9\pm0.4$ \\
MACSJ1149.5+2223	& 0.545	& 1.28	& $9.3^{+0.9}_{-0.8}$	& $47.2\pm1.2$	& $8.5^{+1.1}_{-0.7}$	& $35.3\pm1.1$ \\
A1413	& 0.143	& 1.30	& $7.2^{+0.2}_{-0.2}$	& $17.1\pm0.1$	& $7.1^{+0.3}_{-0.3}$	& $8.6\pm0.1$ \\
CLJ1213+0253	& 0.409	& 0.79	& $3.5^{+0.7}_{-0.6}$	& $2.1\pm0.6$	& $3.9^{+0.9}_{-0.8}$	& $1.6\pm0.5$ \\
RXJ1221+4918	& 0.700	& 0.90	& $6.2^{+0.6}_{-0.6}$	& $12.6\pm0.4$	& $5.9^{+0.7}_{-0.7}$	& $10.1\pm0.4$ \\
CLJ1226.9+3332	& 0.890	& 0.97	& $10.3^{+1.3}_{-0.9}$	& $45.3\pm1.3$	& $10.0^{+1.9}_{-1.4}$	& $24.5\pm1.1$ \\
RXJ1234.2+0947	& 0.229	& 1.15	& $6.6^{+2.3}_{-1.2}$	& $6.2\pm0.4$	& $7.6^{+2.4}_{-2.0}$	& $5.5\pm0.3$ \\
RDCS1252-29	& 1.237	& 0.55	& $4.7^{+0.9}_{-0.7}$	& $6.5\pm0.7$	& $4.6^{+0.9}_{-0.7}$	& $5.4\pm0.7$ \\
A1682	& 0.234	& 1.20	& $6.1^{+1.3}_{-1.0}$	& $12.4\pm1.4$	& $5.8^{+2.0}_{-1.2}$	& $9.7\pm1.2$ \\
MACSJ1311.0-0310	& 0.494	& 0.97	& $5.5^{+0.3}_{-0.2}$	& $17.7\pm0.3$	& $6.5^{+0.8}_{-0.6}$	& $7.2\pm0.2$ \\
A1689	& 0.183	& 1.40	& $9.0^{+0.3}_{-0.3}$	& $39.4\pm0.3$	& $8.4^{+0.4}_{-0.3}$	& $15.7\pm0.2$ \\
RXJ1317.4+2911	& 0.805	& 0.51	& $2.4^{+0.7}_{-0.6}$	& $1.1\pm0.3$	& $2.2^{+0.8}_{-0.4}$	& $0.9\pm0.7$ \\
CLJ1334+5031	& 0.620	& 0.88	& $4.6^{+1.3}_{-1.2}$	& $7.6\pm0.9$	& $5.2^{+2.1}_{-1.5}$	& $5.8\pm0.7$ \\
A1763	& 0.223	& 1.38	& $8.1^{+0.4}_{-0.4}$	& $21.3\pm0.3$	& $8.1^{+0.5}_{-0.5}$	& $14.5\pm0.3$ \\
RXJ1347.5-1145	& 0.451	& 1.51	& $12.6^{+0.4}_{-0.4}$	& $155.6\pm1.4$	& $14.2^{+1.4}_{-1.4}$	& $44.1\pm0.8$ \\
RXJ1350.0+6007	& 0.804	& 0.72	& $4.2^{+0.9}_{-0.7}$	& $5.5\pm0.5$	& $4.5^{+1.0}_{-0.8}$	& $4.6\pm0.5$ \\
CLJ1354-0221	& 0.546	& 0.74	& $3.8^{+0.7}_{-0.6}$	& $3.6\pm0.3$	& $3.1^{+0.9}_{-0.5}$	& $3.0\pm0.3$ \\
CLJ1415.1+3612	& 1.030	& 0.65	& $5.3^{+0.7}_{-0.6}$	& $11.9\pm0.7$	& $4.3^{+0.6}_{-0.6}$	& $7.5\pm0.6$ \\
RXJ1416+4446	& 0.400	& 0.87	& $3.5^{+0.3}_{-0.2}$	& $5.9\pm0.5$	& $3.9^{+0.5}_{-0.4}$	& $3.6\pm0.4$ \\
MACSJ1423.8+2404	& 0.543	& 1.01	& $6.2^{+0.5}_{-0.4}$	& $31.6\pm1.0$	& $6.0^{+1.0}_{-0.8}$	& $12.0\pm0.7$ \\
A1914	& 0.171	& 1.40	& $9.6^{+0.3}_{-0.3}$	& $34.3\pm0.3$	& $8.5^{+0.6}_{-0.4}$	& $14.0\pm0.2$ \\
A1942	& 0.224	& 0.97	& $4.6^{+0.3}_{-0.3}$	& $4.0\pm0.1$	& $4.2^{+0.3}_{-0.3}$	& $3.0\pm0.1$ \\
MS1455.0+2232	& 0.258	& 1.07	& $4.5^{+0.1}_{-0.1}$	& $21.9\pm0.2$	& $4.7^{+0.2}_{-0.2}$	& $6.4\pm0.1$ \\
RXJ1504-0248	& 0.215	& 1.40	& $7.1^{+0.2}_{-0.2}$	& $66.6\pm0.7$	& $9.4^{+1.1}_{-1.0}$	& $15.1\pm0.5$ \\
A2034	& 0.113	& 1.23	& $6.6^{+0.1}_{-0.2}$	& $9.4\pm0.1$	& $6.3^{+0.2}_{-0.2}$	& $6.6\pm0.1$ \\
A2069	& 0.116	& 1.21	& $6.0^{+0.2}_{-0.2}$	& $6.4\pm0.1$	& $5.9^{+0.3}_{-0.3}$	& $5.5\pm0.1$ \\
RXJ1525+0958	& 0.516	& 0.84	& $3.7^{+0.3}_{-0.3}$	& $6.7\pm0.4$	& $3.5^{+0.3}_{-0.4}$	& $5.7\pm0.4$ \\
RXJ1532.9+3021	& 0.345	& 1.14	& $5.1^{+0.2}_{-0.2}$	& $38.0\pm0.9$	& $6.3^{+1.0}_{-0.8}$	& $11.7\pm0.5$ \\
A2111	& 0.229	& 1.23	& $6.4^{+0.6}_{-0.5}$	& $11.4\pm0.3$	& $6.4^{+0.7}_{-0.6}$	& $8.3\pm0.3$ \\
A2125	& 0.246	& 0.79	& $2.5^{+0.1}_{-0.1}$	& $2.0\pm0.1$	& $2.4^{+0.2}_{-0.2}$	& $1.7\pm0.1$ \\
A2163	& 0.203	& 1.91	& $14.7^{+0.8}_{-0.9}$	& $93.9\pm1.3$	& $15.2^{+1.2}_{-1.2}$	& $58.6\pm1.2$ \\
MACSJ1621.3+3810	& 0.463	& 1.03	& $6.1^{+0.3}_{-0.3}$	& $19.8\pm0.6$	& $6.2^{+0.5}_{-0.5}$	& $9.2\pm0.5$ \\
MS1621.5+2640	& 0.426	& 1.04	& $6.1^{+0.6}_{-0.6}$	& $11.2\pm0.4$	& $5.8^{+0.7}_{-0.6}$	& $8.7\pm0.4$ \\
A2204	& 0.152	& 1.44	& $7.1^{+0.2}_{-0.2}$	& $41.1\pm0.4$	& $8.4^{+0.8}_{-0.6}$	& $13.1\pm0.3$ \\
A2218	& 0.176	& 1.21	& $6.4^{+0.2}_{-0.2}$	& $13.3\pm0.1$	& $6.0^{+0.3}_{-0.3}$	& $8.3\pm0.1$ \\
CLJ1641+4001	& 0.464	& 0.77	& $3.5^{+0.4}_{-0.4}$	& $2.7\pm0.3$	& $3.5^{+0.6}_{-0.5}$	& $2.0\pm0.3$ \\
RXJ1701+6414	& 0.453	& 0.88	& $3.8^{+0.3}_{-0.3}$	& $6.8\pm0.6$	& $4.1^{+0.5}_{-0.4}$	& $4.5\pm0.6$ \\
RXJ1716.9+6708	& 0.813	& 0.81	& $6.3^{+1.0}_{-0.8}$	& $13.5\pm0.7$	& $5.7^{+1.1}_{-0.9}$	& $9.0\pm0.7$ \\
A2259	& 0.164	& 1.11	& $5.2^{+0.3}_{-0.3}$	& $8.8\pm0.2$	& $5.2^{+0.4}_{-0.4}$	& $5.3\pm0.2$ \\
RXJ1720.1+2638	& 0.164	& 1.27	& $5.9^{+0.1}_{-0.1}$	& $22.3\pm0.2$	& $6.8^{+0.5}_{-0.3}$	& $8.4\pm0.2$ \\
MACSJ1720.2+3536	& 0.387	& 1.15	& $6.2^{+0.4}_{-0.3}$	& $25.6\pm0.6$	& $7.2^{+0.9}_{-0.8}$	& $11.5\pm0.4$ \\
A2261	& 0.224	& 1.34	& $7.3^{+0.2}_{-0.2}$	& $29.1\pm0.3$	& $7.3^{+0.4}_{-0.4}$	& $14.3\pm0.2$ \\
A2294	& 0.178	& 1.35	& $8.9^{+0.8}_{-0.7}$	& $15.4\pm0.4$	& $8.4^{+1.1}_{-0.8}$	& $9.4\pm0.3$ \\
MACSJ1824.3+4309	& 0.487	& 0.90	& $4.1^{+0.9}_{-0.6}$	& $5.7\pm0.7$	& $4.8^{+1.4}_{-0.9}$	& $4.5\pm0.5$ \\
MACSJ1931.8-2634	& 0.352	& 1.22	& $5.9^{+0.3}_{-0.3}$	& $46.8\pm0.9$	& $6.7^{+1.1}_{-0.7}$	& $14.7\pm0.5$ \\
RXJ2011.3-5725	& 0.279	& 0.89	& $4.0^{+0.2}_{-0.2}$	& $6.9\pm0.3$	& $3.6^{+0.4}_{-0.4}$	& $3.0\pm0.2$ \\
MS2053.7-0449	& 0.583	& 0.76	& $4.0^{+0.4}_{-0.4}$	& $4.9\pm0.3$	& $3.9^{+0.6}_{-0.5}$	& $3.1\pm0.3$ \\
MACSJ2129.4-0741	& 0.594	& 1.12	& $8.9^{+1.1}_{-0.7}$	& $37.2\pm1.2$	& $8.3^{+1.1}_{-1.1}$	& $22.3\pm1.0$ \\
RXJ2129.6+0005	& 0.235	& 1.22	& $5.2^{+0.2}_{-0.2}$	& $21.4\pm0.5$	& $6.2^{+0.6}_{-0.6}$	& $9.8\pm0.4$ \\
A2409	& 0.148	& 1.18	& $5.5^{+0.3}_{-0.2}$	& $11.5\pm0.2$	& $5.7^{+0.4}_{-0.4}$	& $6.7\pm0.2$ \\
MACSJ2228.5+2036	& 0.412	& 1.29	& $8.2^{+0.7}_{-0.6}$	& $33.4\pm0.8$	& $8.6^{+1.4}_{-0.8}$	& $22.2\pm0.7$ \\
MACSJ2229.7-2755	& 0.324	& 1.03	& $4.1^{+0.2}_{-0.2}$	& $20.5\pm0.8$	& $5.0^{+0.9}_{-0.7}$	& $6.4\pm0.4$ \\
MACSJ2245.0+2637	& 0.301	& 1.05	& $5.2^{+0.3}_{-0.3}$	& $17.1\pm0.5$	& $4.9^{+0.5}_{-0.5}$	& $6.8\pm0.4$ \\
RXJ2247+0337	& 0.200	& 0.67	& $2.7^{+0.7}_{-0.5}$	& $0.4\pm0.1$	& $2.9^{+0.9}_{-0.6}$	& $0.3\pm0.1$ \\
AS1063	& 0.348	& 1.56	& $11.5^{+0.6}_{-0.6}$	& $93.6\pm1.1$	& $11.2^{+1.1}_{-0.9}$	& $42.1\pm0.9$ \\
CLJ2302.8+0844	& 0.722	& 0.76	& $5.4^{+1.5}_{-1.0}$	& $5.2\pm0.4$	& $5.5^{+2.4}_{-1.4}$	& $3.8\pm0.4$ \\
A2631	& 0.273	& 1.26	& $7.0^{+0.7}_{-0.5}$	& $19.5\pm0.5$	& $6.9^{+0.8}_{-0.5}$	& $14.1\pm0.5$ \\

\end{longtable}
\twocolumn
\normalsize

\noindent
curve cleaning was performed by hand to
remove periods of high background that were not detected in the M08
analysis. This generally led to a small decrease in cluster
temperature, consistent with the removal of a harder spectral
component. For two clusters (RXJ1701+6414, RXJ1525+0958) the decrease
was significant compared to the statistical uncertainties on kT (both
decreased from $\approx5\keV$ to $\approx4\keV$). For two other
clusters the change in kT was large, but within the statistical
uncertainties: CLJ0216-1747 decreased from $\approx8\keV$ to
$\approx6\keV$ and CLJ1334+5031 decreased from $\approx16\keV$ to
$\approx6\keV$. For one cluster (CLJ1216+2633), the good time
remaining after the new cleaning was too short for useful analysis,
and this cluster was dropped from the sample, reducing the total
number to 114 clusters. Finally, for AS1063, a redshift of 0.252 was
erroneously used in M08, and this was corrected to 0.348 for the
current analysis.

There have been several significant updates to the \Chandra\
calibration since the analysis presented in M08, which was based on
CALDB version 3.2.3. The most significant changes for the measurement
of cluster temperatures and luminosities were updates to the mirror
effective area, and the ACIS contamination model. \citet{ree10}
recently examined the effect of \Chandra\ calibration changes on the
temperatures of galaxy clusters, finding that on average, temperatures
measured with CALDB 3.1 were $6\%$ higher than those measured with
CALDB 4.2. The average temperature change (for kT measured within \rf)
from M08 to the current analysis was
$<kT_{old}/kT_{new}>=1.00\pm0.06$, where the uncertainty is the
standard deviation of the clusters, and those clusters with
significant individual changes discussed above were excluded. This is
smaller than the change found by \citet{ree10}, but we note that
unlike that work, our study was not optimised to study the effect of
calibration changes; there were numerous other changes between the M08
analysis and the current work, as discussed above.

The bolometric luminosities measured within \rf\ were also compared
with M08, and the new luminosities were found to be higher by $8\%$ on
average ($<L_{X,old}/L_{X,new}>=0.93\pm0.06$). This is consistent with
the $\approx9\%$ decrease in the effective area at low energies
introduced in CALDB 4.1.1\footnote{
  \url{http://asc.harvard.edu/ciao/why/caldb4.1.1_hrma.html}}.

We note that all of the main results in the following sections were
also present with similar statistical significance in the data as
presented in M08, so our results are not sensitive to the various
calibration and analysis changes.

\begin{figure}
\begin{center}
\scalebox{0.34}{\includegraphics*[angle=270]{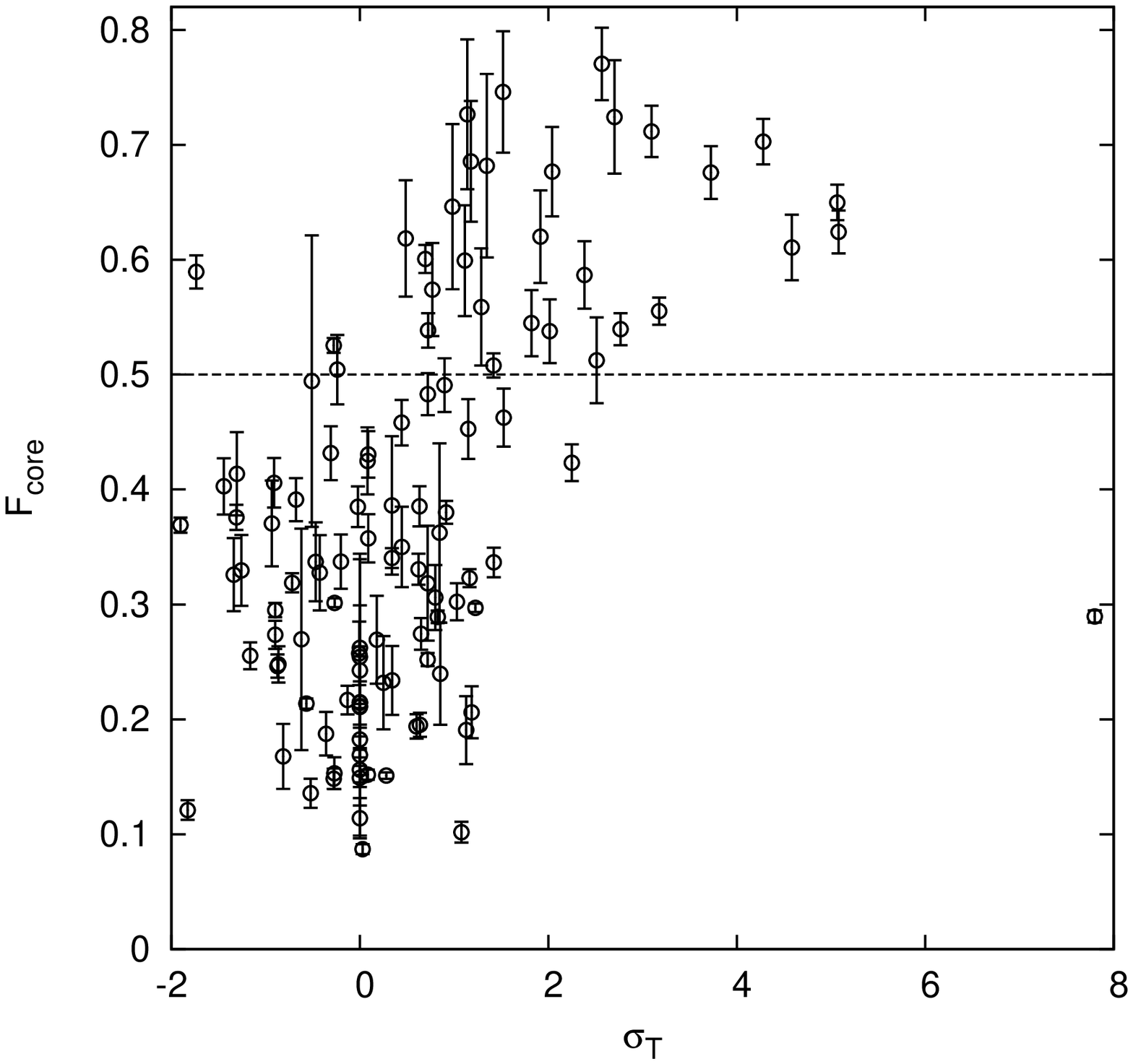}} \\
\scalebox{0.34}{\includegraphics*[angle=270]{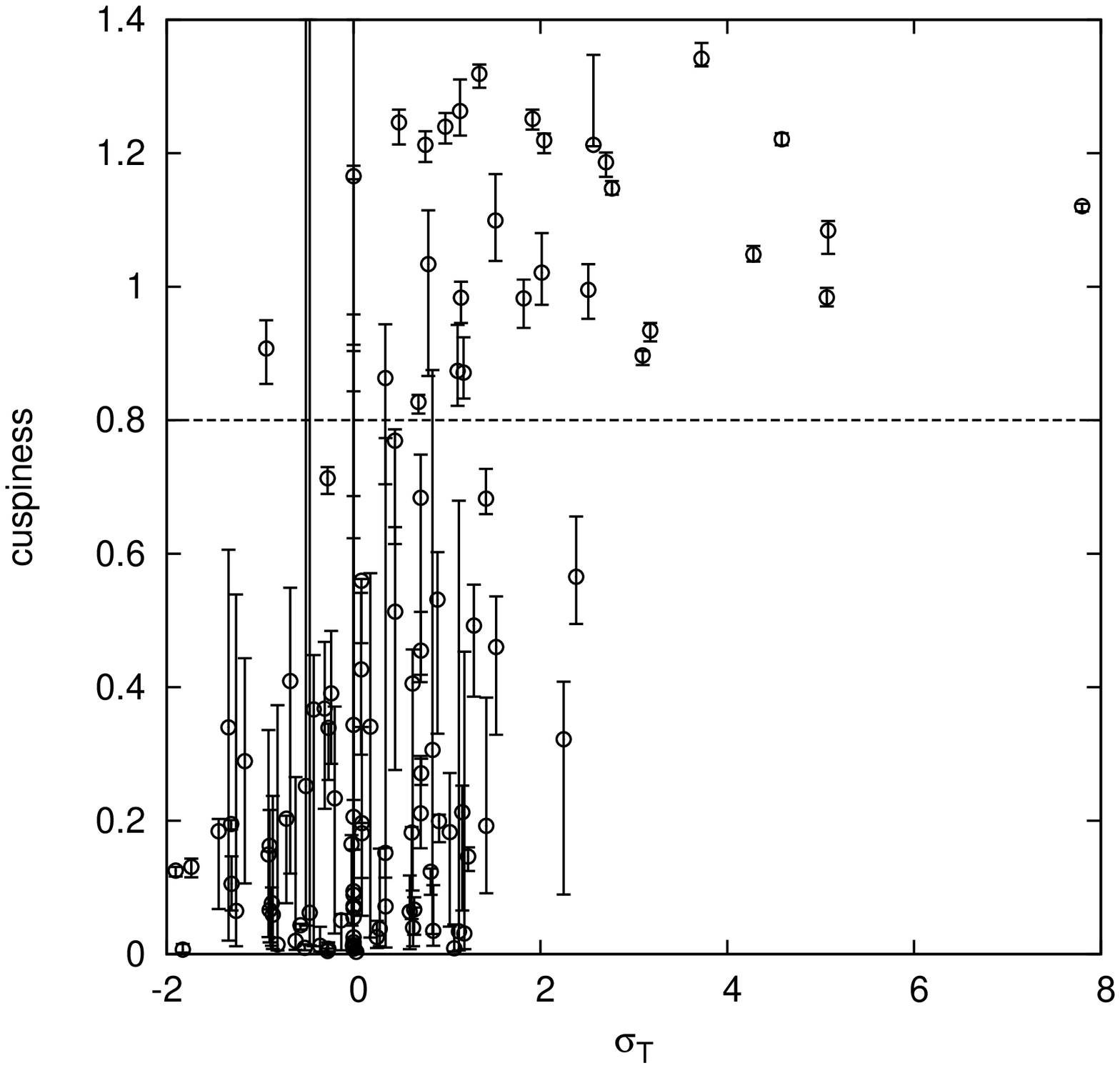}} \\
\caption[]{\label{f.cc} Comparison of simple cool core proxies \Fcore\
  (top panel) and cuspiness (bottom panel) with \sigT. The dashed
  horizontal line in each plot indicates the value of each property
  used to divide the sample into CC and NCC clusters. The outlier at
  $\sigT\approx8$ is the merging, double cool core cluster A115.}
\end{center}
\end{figure}

\begin{figure*}
\begin{center}
\scalebox{0.28}{\includegraphics*[angle=270]{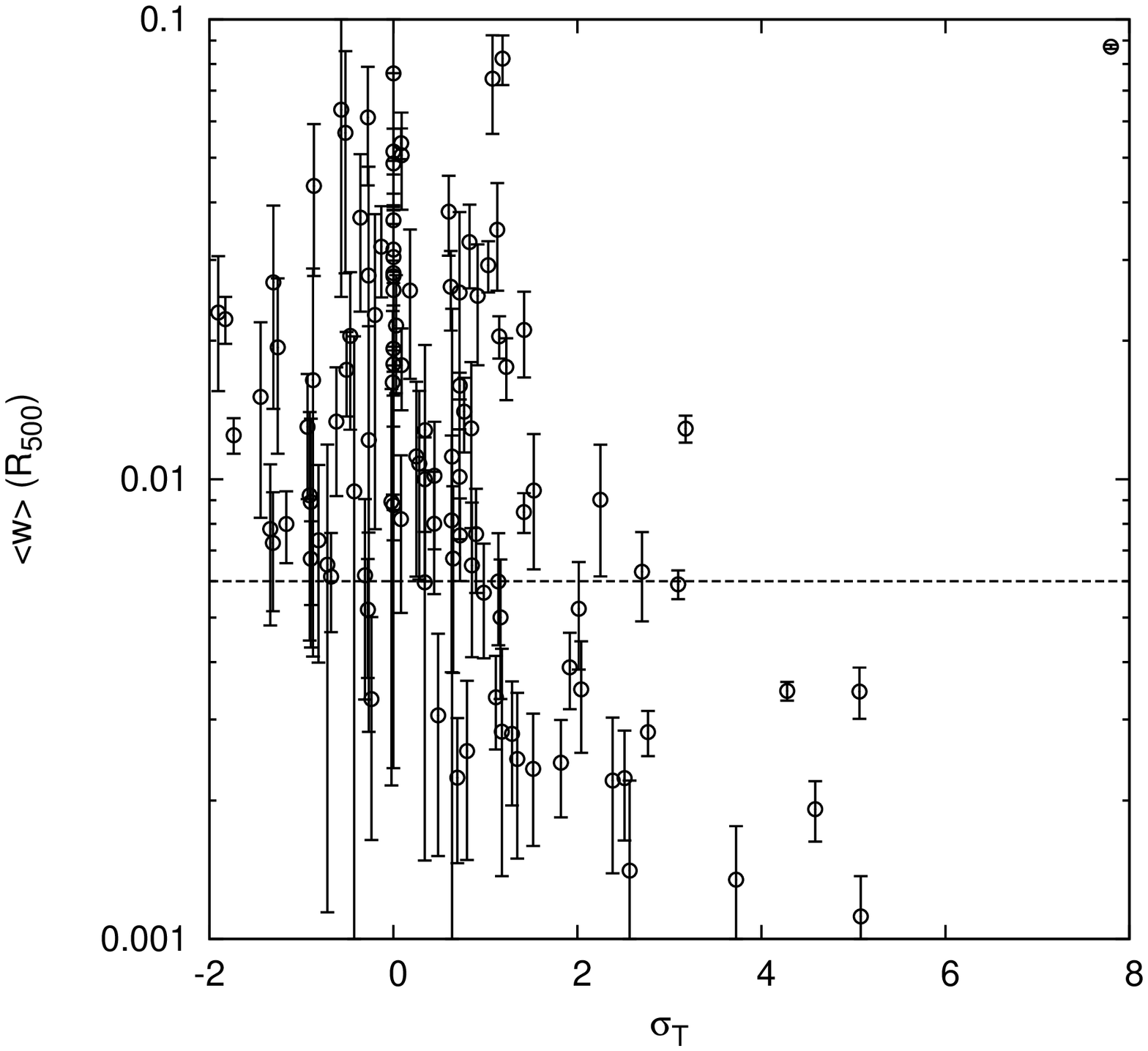}}
\hspace{-2cm}
\scalebox{0.28}{\includegraphics*[angle=270]{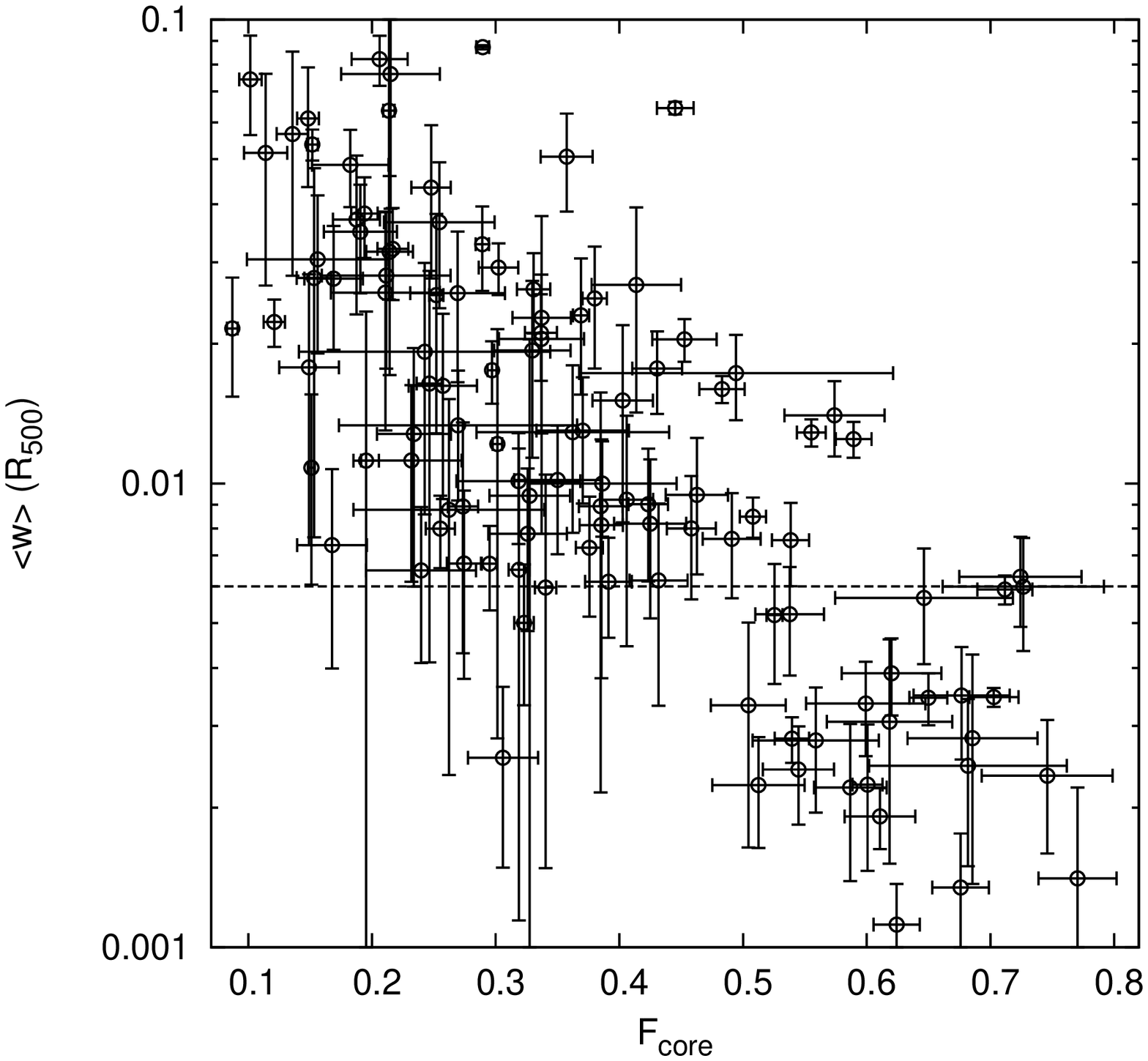}}
\hspace{-2cm}
\scalebox{0.28}{\includegraphics*[angle=270]{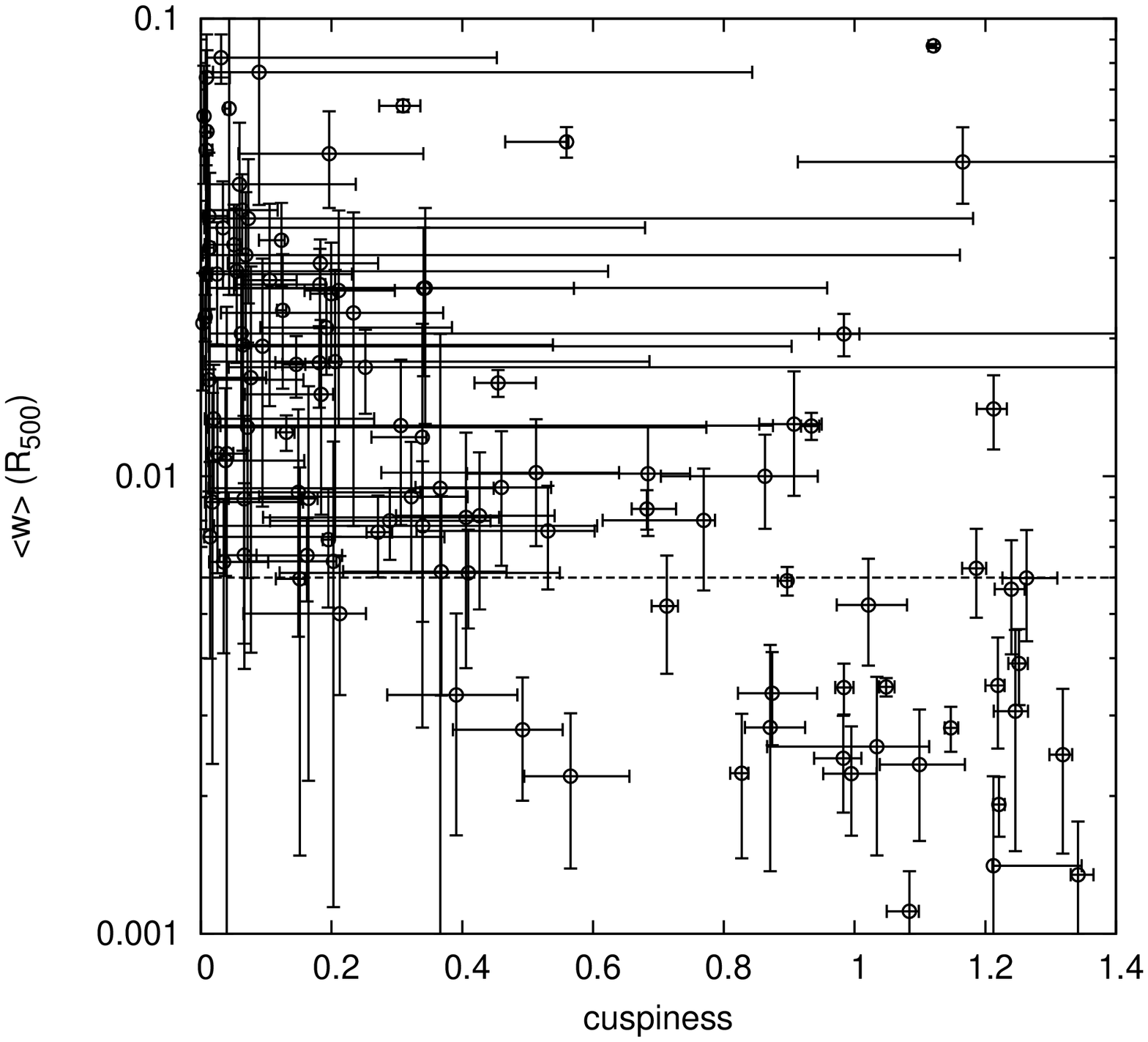}} \\
\caption[]{\label{f.wcc} Comparison of cluster dynamical state with
  cool core proxy. The centroid shift is plotted against \sigT\ (left
  panel), \Fcore\ (centre panel) and cuspiness (right panel). The
  dashed line indicates the value of $\cshift=0.006\rf$ used to
  separate relaxed and unrelaxed clusters.}
\end{center}
\end{figure*}

\section{Quantifying dynamical state and cool core presence}
The dynamical state of the cluster was quantified by measuring the
centroid shift, following the method of \citet{poo06}. The centroid
shift ($\cshift$) was defined as the standard deviation of the
distance between the X-ray peak and centroid, with the latter measured
in a series of circular apertures centred on the X-ray peak and
decreasing in steps of $5\%$ from \rf\ to $0.05\rf$. For some
clusters, \rf\ extended beyond the edges of the detector in some
directions. A maximum radius (R$_w$) within which \cshift\ could be
measured was thus defined for each cluster, such that the maximum
fraction of the area within this radius that was lost to excluded
regions, or off the edge of the detector was $2\%$. Following
\citet{poo06}, the central $30\kpc$ were excluded from these centroid
measurements to increase the sensitivity to faint structure. In an
improvement to M08, the uncertainties on the centroid shifts were
calculated from Monte Carlo randomisations of the X-ray images, with
the measurement of $\cshift$ being repeated for 100 realisations of
the input image with pixels randomised under a Poisson
distribution. This takes into account the effect of the image noise,
while in M08, the error was derived simply from the number of values
used in the standard deviation calculation.

In addition to quantifying cluster morphology with the centroid shift,
two quantities were used to determine if a cluster was likely to host
a cool core. In a recent comprehensive study, \citet{hud10} tested 16
cool core probes, and concluded that for high quality data, a direct
measurement of the central cooling time is preferred, but in lower
quality data the cuspiness of the gas density profile was recommended
\citep[see also][]{vik06c}. As our sample spans a wide range of
redshift and \Lx, cuspiness is a more appropriate choice as it can be
measured reliably for all clusters. Cuspiness is defined as the
logarithmic slope of the gas density profile at a radius of $0.04\rf$,
and was measured using the best fitting gas density models, with the
uncertainty derived from the cuspiness measured for each of the Monte
Carlo randomisations of the density profile. The second quantity used
was the core flux ratio ($\Fcore$), defined as the ratio of the
bolometric unabsorbed flux from the central $0.15\rf$ to the total
within $\rf$. This quantity is the most robust to variations in the
X-ray data quality and angular scale with redshift. \Fcore\ values
were also computed using rest-frame $(0.5-2)\keV$ band unabsorbed
fluxes, but this made no significant difference, so bolometric fluxes
were used.

The most direct test of the presence of a cool core would be a
temperature profile, but the data available are not sufficient for
full temperature profiles for the majority of clusters in our
sample. Instead, the projected temperature was measured for each
cluster in a core ($0-0.15\rf$) aperture and outer ($0.15-0.3\rf$)
aperture, chosen to sample respectively the central dip and subsequent
peak of the temperature profile of a typical cool core cluster
\citep{vik06a}. The significance of the difference between the outer
and core temperatures then gives a direct measurement of any cool
core. This quantity ($\sigT$) was defined as the difference in
temperatures divided by the quadrature sum of their measurement
errors, such that positive $\sigT$ indicates a cooler core
region. This probe is sensitive to the data quality, and not just the
strength of any cool core, so cool core clusters with low signal to
noise observations could have $\sigT\approx0$. However, large values
of $\sigT$ are an unambiguous indicator of cool core presence and can
be used to calibrate other cool core proxies which are less sensitive
to data quality. The dynamical properties of the clusters are given in
Table \ref{t.cc}.

Figure \ref{f.cc} shows the cool core proxies plotted against $\sigT$,
and these plots were used to define the value of each proxy used to
split the sample into clusters with and without cool cores (referred
to as CC and NCC clusters hereafter). Almost all clusters with
$\sigT\ga2$ have $\Fcore>0.5$ (i.e. more than half of the flux within
$\rf$ comes from the central $15\%$), and we thus define $\Fcore=0.5$
as the border between CC and NCC clusters in the $\Fcore$ parameter
space, giving 31 CC and 83 NCC clusters. Similarly, almost all
clusters with $\sigT\ga2$ have cuspiness $>0.8$, so this is defined as
the CC/NCC boundary in cuspiness, giving 30 CC and 84 NCC clusters
\citep[this is similar to the value of 0.7 adopted to define strong
cool cores in][]{vik06c}.

There is substantial overlap between the subsamples defined in
$\Fcore$ and cuspiness, and unless specified otherwise, in the
following sections all results are shown for the CC/NCC populations
defined by $\Fcore$.

The dynamical state of the clusters is also related to their cool core
status in the sense that cool core clusters tend to be dynamically
relaxed. This is illustrated in figure \ref{f.wcc} which shows the
centroid shift parameter plotted against each of the cool core
measures. The most relaxed clusters show evidence for cool cores in
all of the different measures (albeit with a larger dispersion for
cuspiness), and we adopt a threshold of $\cshift=0.006\rf$ below which
clusters are classed as ``relaxed'' (28 clusters), and above which
as ``unrelaxed'' (86 clusters). This a minor update to the value
of $\cshift=0.005\rf$ used to segregate the clusters in
\citet{mau07b}.

We finally define a split into 21 clusters that passed the combined
filter of all three of the above tests ($\Fcore>0.5$, cuspiness
$>0.8$, $\cshift>0.006\rf$), and 93 clusters that do not. These
clusters that are both relaxed and pass both cool core filters are
referred to as RCC clusters (for ``relaxed cool core clusters''),
while the complementary set are referred to as NRCC clusters.

\section{Strong self-similarity}
\begin{figure}
\begin{center}
\scalebox{0.32}{\includegraphics*[angle=270]{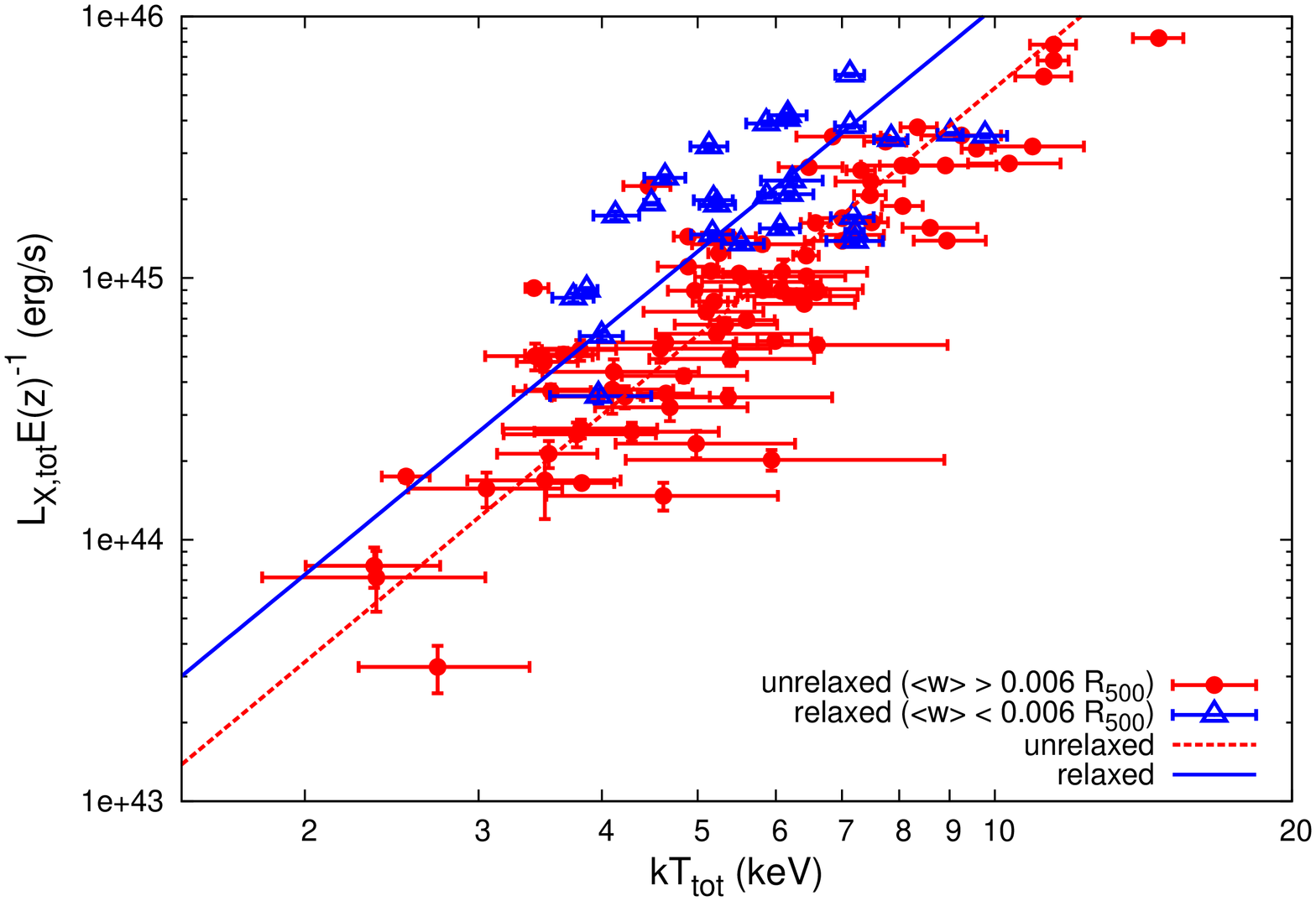}}\\
\scalebox{0.32}{\includegraphics*[angle=270]{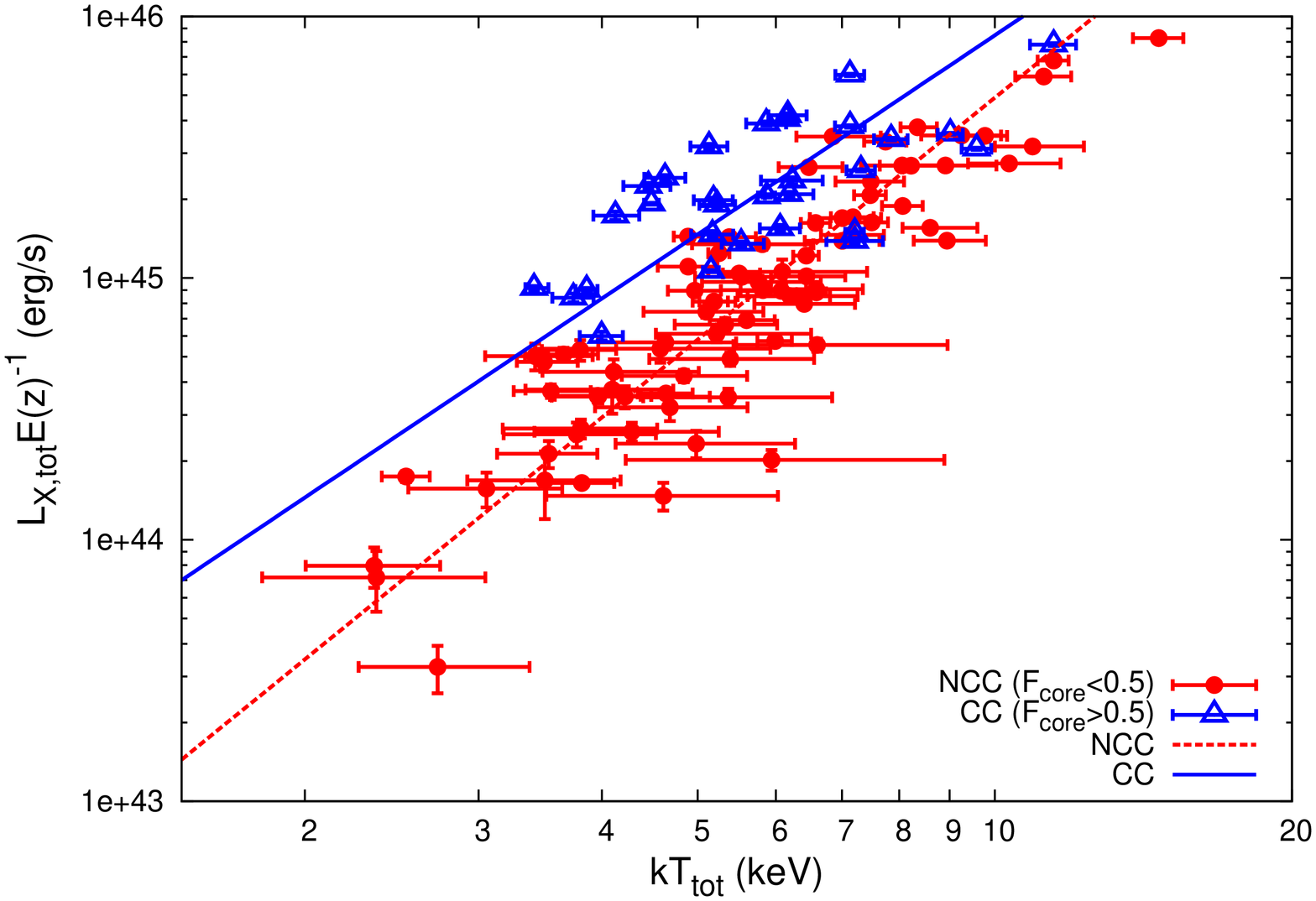}} \\
\caption[]{\label{f.ltc} \LT\ relation for the cluster sample, separated
  into relaxed/unrelaxed subsamples (top panel) or CC/NCC subsamples
  (bottom panel). \Lx\
  and kT were measured within \rf, including the core
  regions. Luminosities are corrected for self-similar evolution as
  indicated on the ordinate axis. The lines show the best fitting
  power laws determined from a BCES orthogonal regression fit (see text
for details).}
\end{center}
\end{figure}

The strong self-similarity of the \LT\ relation of the population was
investigated. Figure \ref{f.ltc} shows the \LT\ relation where both
quantities were measured within \rf, including the core regions. A
power law of the form
\begin{eqnarray}
  \label{eq.lt}
  \Lx & = & E(z)A_{LT}(kT/T_*)^{B_{LT}},
\end{eqnarray}
was fit to the data, with $T_*$ set at $6\keV$ (close to the median
temperature of $5.9\keV$ for the full sample). The model was fit in
log space using BCES orthogonal regression \citep{akr96}, and the
intrinsic scatter about the best fitting model was estimated by
adding an extra error term in quadrature to every data point until the
reduced $\chisq$ was unity \citep[see][for details]{mau07b}. The
intrinsic scatter of a population is thus measured in the \Lx\
direction about the best fitting \LT\ relation for that

\small
\onecolumn
\begin{longtable}{lrccll}
  \caption{Summary of the cluster dynamical properties. \sigT\
    indicates the significance of the temperature decrease in the core
    region, \Fcore\ is the ratio of the bolometric flux within
    $0.15\rf$ to that within $\rf$, cuspiness is the logarithmic slope
    of the gas density profile at $0.04\rf$, and \cshift\ is the
    centroid shift, with R$_w$ giving the maximum radius
    within which \cshift\ could be measured.}\label{t.cc}\\
\hline
Cluster & \sigT & \Fcore & cuspiness & \cshift ($10^{-3}\rf$) & R$_w$ ($\rf$) \\ \hline
\endfirsthead

\caption{\emph{continued}}\\
\hline
Cluster & \sigT & \Fcore & cuspiness & \cshift ($10^{-3}\rf$) & R$_w$ ($\rf$) \\ \hline
\endhead
\hline
\multicolumn{5}{r}{\emph{continued on next page}}
\endfoot
\hline
\endlastfoot

MS0015.9+1609	& -0.9	& $0.298\pm0.006$	& $0.162^{+0.054}_{-0.095}$	& $6.7\pm1.4$	& $1.00$ \\
RXJ0027.6+2616	& -0.0	& $0.240\pm0.028$	& $0.013^{+0.144}_{-0.008}$	& $16.2\pm7.0$	& $1.00$ \\
CLJ0030+2618	& 0.0	& $0.262\pm0.081$	& $0.018^{+0.138}_{-0.011}$	& $8.8\pm6.4$	& $1.00$ \\
A68	& -0.9	& $0.410\pm0.022$	& $0.149^{+0.186}_{-0.123}$	& $9.2\pm4.8$	& $1.00$ \\
A115	& 7.8	& $0.292\pm0.005$	& $1.121^{+0.004}_{-0.008}$	& $87.2\pm0.9$	& $1.00$ \\
A209	& 1.2	& $0.321\pm0.008$	& $0.213^{+0.040}_{-0.147}$	& $5.0\pm1.7$	& $0.95$ \\
CLJ0152.7-1357S	& 1.2	& $0.212\pm0.022$	& $0.031^{+0.422}_{-0.024}$	& $82.2\pm10.2$	& $1.00$ \\
A267	& -1.3	& $0.415\pm0.036$	& $0.105^{+0.041}_{-0.040}$	& $26.8\pm12.6$	& $1.00$ \\
CLJ0152.7-1357N	& -0.4	& $0.192\pm0.018$	& $0.012^{+0.029}_{-0.009}$	& $37.0\pm13.9$	& $1.00$ \\
MACSJ0159.8-0849	& 2.8	& $0.547\pm0.014$	& $1.147^{+0.011}_{-0.010}$	& $2.8\pm0.3$	& $1.00$ \\
CLJ0216-1747	& 0.0	& $0.255\pm0.038$	& $0.073^{+1.109}_{-0.068}$	& $36.6\pm12.7$	& $1.00$ \\
RXJ0232.2-4420	& 1.1	& $0.455\pm0.027$	& $0.984^{+0.024}_{-0.038}$	& $20.4\pm2.2$	& $1.00$ \\
MACSJ0242.5-2132	& 1.1	& $0.726\pm0.065$	& $1.263^{+0.047}_{-0.037}$	& $6.0\pm1.6$	& $1.00$ \\
A383	& 4.6	& $0.616\pm0.029$	& $1.221^{+0.009}_{-0.009}$	& $1.9\pm0.3$	& $1.00$ \\
MACSJ0257.6-2209	& 0.9	& $0.496\pm0.024$	& $0.531^{+0.071}_{-0.201}$	& $7.6\pm1.9$	& $1.00$ \\
MS0302.7+1658	& -0.5	& $0.487\pm0.124$	& $0.252^{+3.511}_{-0.209}$	& $17.3\pm3.6$	& $1.00$ \\
CLJ0318-0302	& -0.5	& $0.339\pm0.034$	& $0.062^{+3.450}_{-0.049}$	& $20.5\pm7.7$	& $1.00$ \\
MACSJ0329.6-0211	& 0.8	& $0.593\pm0.042$	& $1.213^{+0.020}_{-0.026}$	& $14.0\pm2.6$	& $1.00$ \\
MACSJ0404.6+1109	& 0.6	& $0.194\pm0.011$	& $0.063^{+0.054}_{-0.056}$	& $38.2\pm7.5$	& $1.00$ \\
MACSJ0429.6-0253	& 1.9	& $0.621\pm0.040$	& $1.252^{+0.014}_{-0.016}$	& $3.9\pm0.7$	& $1.00$ \\
RXJ0439.0+0715	& 0.4	& $0.460\pm0.020$	& $0.769^{+0.017}_{-0.155}$	& $8.0\pm2.4$	& $1.00$ \\
RXJ0439+0520	& 1.0	& $0.648\pm0.073$	& $1.240^{+0.020}_{-0.025}$	& $5.7\pm1.6$	& $1.00$ \\
MACSJ0451.9+0006	& -1.3	& $0.345\pm0.030$	& $0.065^{+0.474}_{-0.053}$	& $19.3\pm8.0$	& $1.00$ \\
A521	& 0.1	& $0.154\pm0.004$	& $0.559^{+0.003}_{-0.094}$	& $53.8\pm4.1$	& $1.00$ \\
A520	& -0.6	& $0.213\pm0.005$	& $0.043^{+0.002}_{-0.006}$	& $63.6\pm38.7$	& $1.00$ \\
MS0451.6-0305	& -0.2	& $0.332\pm0.023$	& $0.233^{+0.137}_{-0.203}$	& $22.7\pm15.0$	& $1.00$ \\
CLJ0522-3625	& 0.3	& $0.253\pm0.033$	& $0.071^{+0.702}_{-0.062}$	& $12.8\pm6.8$	& $1.00$ \\
CLJ0542.8-4100	& -0.9	& $0.242\pm0.016$	& $0.059^{+0.178}_{-0.051}$	& $43.4\pm15.8$	& $1.00$ \\
MACSJ0647.7+7015	& -0.3	& $0.437\pm0.021$	& $0.368^{+0.100}_{-0.150}$	& $6.2\pm2.9$	& $1.00$ \\
1E0657-56	& 1.2	& $0.301\pm0.004$	& $0.146^{+0.014}_{-0.022}$	& $17.5\pm2.7$	& $1.00$ \\
MACSJ0717.5+3745	& 0.7	& $0.256\pm0.006$	& $0.211^{+0.086}_{-0.052}$	& $25.5\pm12.6$	& $1.00$ \\
A586	& -0.2	& $0.502\pm0.030$	& $0.391^{+0.093}_{-0.106}$	& $3.3\pm1.7$	& $0.75$ \\
MACSJ0744.9+3927	& 0.7	& $0.487\pm0.019$	& $0.455^{+0.058}_{-0.036}$	& $16.0\pm1.1$	& $1.00$ \\
A665	& 0.8	& $0.288\pm0.005$	& $0.123^{+0.005}_{-0.034}$	& $32.8\pm6.8$	& $0.75$ \\
A697	& 0.3	& $0.347\pm0.009$	& $0.151^{+0.008}_{-0.037}$	& $6.0\pm4.5$	& $0.95$ \\
CLJ0848.7+4456	& 0.0	& $0.212\pm0.057$	& $0.055^{+0.568}_{-0.047}$	& $28.0\pm10.4$	& $1.00$ \\
ZWCLJ1953	& 0.1	& $0.433\pm0.020$	& $0.181^{+0.015}_{-0.067}$	& $17.7\pm3.6$	& $1.00$ \\
CLJ0853+5759	& 0.0	& $0.120\pm0.022$	& $0.008^{+0.010}_{-0.004}$	& $51.6\pm24.9$	& $1.00$ \\
MS0906.5+1110	& -2.3	& $0.443\pm0.015$	& $0.309^{+0.027}_{-0.036}$	& $64.5\pm2.0$	& $0.90$ \\
RXJ0910+5422	& 0.0	& $0.196\pm0.042$	& $0.343^{+0.615}_{-0.333}$	& $25.8\pm12.8$	& $1.00$ \\
A773	& -1.3	& $0.378\pm0.011$	& $0.195^{+0.006}_{-0.008}$	& $7.3\pm2.1$	& $1.00$ \\
A781	& -0.3	& $0.154\pm0.009$	& $0.005^{+0.004}_{-0.002}$	& $61.2\pm17.7$	& $1.00$ \\
CLJ0926+1242	& 0.8	& $0.303\pm0.032$	& $1.034^{+0.080}_{-0.168}$	& $2.6\pm1.1$	& $1.00$ \\
RBS797	& 1.5	& $0.743\pm0.054$	& $1.099^{+0.069}_{-0.061}$	& $2.3\pm0.7$	& $1.00$ \\
MACSJ0949.8+1708	& 0.6	& $0.393\pm0.018$	& $0.406^{+0.051}_{-0.310}$	& $8.1\pm4.3$	& $1.00$ \\
CLJ0956+4107	& 0.0	& $0.219\pm0.020$	& $0.013^{+0.006}_{-0.006}$	& $31.6\pm14.5$	& $1.00$ \\
A907	& 3.2	& $0.555\pm0.012$	& $0.934^{+0.012}_{-0.016}$	& $12.9\pm0.9$	& $0.95$ \\
MS1006.0+1202	& 1.4	& $0.340\pm0.013$	& $0.192^{+0.192}_{-0.101}$	& $21.1\pm4.5$	& $1.00$ \\
MS1008.1-1224	& 0.1	& $0.358\pm0.021$	& $0.196^{+0.144}_{-0.139}$	& $50.6\pm12.1$	& $1.00$ \\
ZW3146	& 5.1	& $0.654\pm0.016$	& $0.984^{+0.014}_{-0.013}$	& $3.5\pm0.4$	& $1.00$ \\
CLJ1113.1-2615	& 0.8	& $0.355\pm0.077$	& $0.306^{+0.570}_{-0.276}$	& $12.9\pm5.1$	& $1.00$ \\
A1204	& 2.7	& $0.727\pm0.051$	& $1.186^{+0.015}_{-0.022}$	& $6.3\pm1.4$	& $1.00$ \\
CLJ1117+1745	& 0.0	& $0.134\pm0.028$	& $0.205^{+0.481}_{-0.199}$	& $17.8\pm10.4$	& $1.00$ \\
CLJ1120+4318	& -0.4	& $0.352\pm0.036$	& $0.366^{+0.082}_{-0.354}$	& $9.4\pm11.0$	& $1.00$ \\
RXJ1121+2327	& -0.5	& $0.146\pm0.013$	& $0.009^{+0.006}_{-0.004}$	& $56.7\pm28.6$	& $1.00$ \\
A1240	& 0.0	& $0.091\pm0.005$	& $0.003^{+0.001}_{-0.001}$	& $21.6\pm6.2$	& $1.00$ \\
MACSJ1131.8-1955	& 1.0	& $0.306\pm0.016$	& $0.183^{+0.088}_{-0.142}$	& $29.2\pm3.7$	& $1.00$ \\
MS1137.5+6625	& 0.1	& $0.435\pm0.030$	& $0.426^{+0.115}_{-0.128}$	& $8.2\pm3.1$	& $1.00$ \\
MACSJ1149.5+2223	& -0.9	& $0.251\pm0.010$	& $0.076^{+0.024}_{-0.064}$	& $16.4\pm12.3$	& $1.00$ \\
A1413	& -0.3	& $0.498\pm0.006$	& $0.713^{+0.017}_{-0.023}$	& $5.2\pm1.5$	& $0.85$ \\
CLJ1213+0253	& 0.0	& $0.258\pm0.118$	& $0.094^{+0.809}_{-0.082}$	& $19.2\pm10.7$	& $1.00$ \\
RXJ1221+4918	& 0.6	& $0.195\pm0.010$	& $0.040^{+0.030}_{-0.028}$	& $11.2\pm12.3$	& $1.00$ \\
CLJ1226.9+3332	& 1.5	& $0.459\pm0.028$	& $0.460^{+0.076}_{-0.132}$	& $9.5\pm3.1$	& $1.00$ \\
RXJ1234.2+0947	& 1.1	& $0.110\pm0.009$	& $0.009^{+0.036}_{-0.005}$	& $74.3\pm18.0$	& $1.00$ \\
RDCS1252-29	& -0.8	& $0.168\pm0.029$	& $0.014^{+0.359}_{-0.009}$	& $7.4\pm3.4$	& $1.00$ \\
A1682	& 0.2	& $0.222\pm0.039$	& $0.026^{+0.024}_{-0.017}$	& $11.2\pm5.1$	& $0.80$ \\
MACSJ1311.0-0310	& 2.4	& $0.594\pm0.029$	& $0.565^{+0.090}_{-0.071}$	& $2.2\pm0.8$	& $1.00$ \\
A1689	& 0.7	& $0.600\pm0.012$	& $0.827^{+0.010}_{-0.018}$	& $2.2\pm0.8$	& $1.00$ \\
RXJ1317.4+2911	& 0.0	& $0.174\pm0.120$	& $0.069^{+1.092}_{-0.066}$	& $30.4\pm11.4$	& $1.00$ \\
CLJ1334+5031	& 0.0	& $0.226\pm0.040$	& $0.089^{+0.754}_{-0.070}$	& $76.3\pm37.2$	& $1.00$ \\
A1763	& -0.7	& $0.319\pm0.008$	& $0.203^{+0.005}_{-0.127}$	& $6.5\pm5.4$	& $0.90$ \\
RXJ1347.5-1145	& 3.1	& $0.717\pm0.023$	& $0.897^{+0.007}_{-0.014}$	& $5.9\pm0.4$	& $1.00$ \\
RXJ1350.0+6007	& 1.1	& $0.161\pm0.024$	& $0.034^{+0.646}_{-0.028}$	& $34.9\pm9.2$	& $1.00$ \\
CLJ1354-0221	& 0.0	& $0.157\pm0.021$	& $0.025^{+0.206}_{-0.018}$	& $27.6\pm8.2$	& $1.00$ \\
CLJ1415.1+3612	& -0.9	& $0.369\pm0.038$	& $0.907^{+0.042}_{-0.053}$	& $13.0\pm3.9$	& $1.00$ \\
RXJ1416+4446	& 0.3	& $0.394\pm0.061$	& $0.863^{+0.081}_{-0.159}$	& $10.0\pm2.3$	& $1.00$ \\
MACSJ1423.8+2404	& 0.5	& $0.621\pm0.051$	& $1.246^{+0.019}_{-0.033}$	& $3.1\pm1.5$	& $1.00$ \\
A1914	& -1.7	& $0.592\pm0.015$	& $0.131^{+0.013}_{-0.016}$	& $12.5\pm1.1$	& $0.65$ \\
A1942	& -1.2	& $0.256\pm0.012$	& $0.289^{+0.154}_{-0.183}$	& $8.0\pm1.4$	& $1.00$ \\
MS1455.0+2232	& 4.3	& $0.709\pm0.020$	& $1.048^{+0.013}_{-0.011}$	& $3.5\pm0.2$	& $1.00$ \\
RXJ1504-0248	& 2.6	& $0.774\pm0.037$	& $1.212^{+0.135}_{-0.002}$	& $1.4\pm0.8$	& $1.00$ \\
A2034	& -0.3	& $0.302\pm0.004$	& $0.339^{+0.006}_{-0.078}$	& $12.2\pm9.3$	& $0.75$ \\
A2069	& 0.3	& $0.147\pm0.003$	& $0.038^{+0.120}_{-0.026}$	& $10.8\pm4.8$	& $0.75$ \\
RXJ1525+0958	& -0.3	& $0.162\pm0.015$	& $0.008^{+0.009}_{-0.004}$	& $27.7\pm20.1$	& $1.00$ \\
RXJ1532.9+3021	& 1.2	& $0.691\pm0.054$	& $0.871^{+0.053}_{-0.039}$	& $2.8\pm1.5$	& $1.00$ \\
A2111	& 0.6	& $0.274\pm0.014$	& $0.067^{+0.019}_{-0.038}$	& $6.7\pm2.9$	& $1.00$ \\
A2125	& -1.8	& $0.125\pm0.009$	& $0.007^{+0.008}_{-0.005}$	& $22.3\pm2.6$	& $1.00$ \\
A2163	& 0.9	& $0.376\pm0.010$	& $0.199^{+0.010}_{-0.032}$	& $25.0\pm7.4$	& $0.60$ \\
MACSJ1621.3+3810	& 2.5	& $0.538\pm0.038$	& $0.995^{+0.039}_{-0.043}$	& $2.2\pm0.6$	& $1.00$ \\
MS1621.5+2640	& -0.1	& $0.226\pm0.013$	& $0.051^{+0.009}_{-0.045}$	& $32.1\pm7.2$	& $1.00$ \\
A2204	& 3.7	& $0.681\pm0.023$	& $1.342^{+0.024}_{-0.011}$	& $1.3\pm0.4$	& $0.75$ \\
A2218	& -1.9	& $0.373\pm0.007$	& $0.125^{+0.005}_{-0.009}$	& $23.0\pm7.5$	& $1.00$ \\
CLJ1641+4001	& 0.9	& $0.261\pm0.051$	& $0.035^{+0.068}_{-0.022}$	& $6.5\pm2.4$	& $1.00$ \\
RXJ1701+6414	& 0.7	& $0.328\pm0.052$	& $0.684^{+0.064}_{-0.276}$	& $10.1\pm2.7$	& $1.00$ \\
RXJ1716.9+6708	& -1.3	& $0.335\pm0.033$	& $0.340^{+0.267}_{-0.319}$	& $7.8\pm3.0$	& $1.00$ \\
A2259	& -0.0	& $0.394\pm0.018$	& $0.165^{+0.014}_{-0.122}$	& $8.9\pm6.8$	& $0.85$ \\
RXJ1720.1+2638	& 5.1	& $0.625\pm0.018$	& $1.084^{+0.014}_{-0.035}$	& $1.1\pm0.3$	& $1.00$ \\
MACSJ1720.2+3536	& 1.8	& $0.552\pm0.029$	& $0.983^{+0.028}_{-0.044}$	& $2.4\pm0.6$	& $1.00$ \\
A2261	& 1.4	& $0.509\pm0.011$	& $0.682^{+0.044}_{-0.023}$	& $8.5\pm0.8$	& $1.00$ \\
A2294	& -0.7	& $0.388\pm0.019$	& $0.409^{+0.140}_{-0.289}$	& $6.1\pm1.5$	& $0.65$ \\
MACSJ1824.3+4309	& 0.0	& $0.210\pm0.036$	& $1.166^{+0.246}_{-0.252}$	& $48.6\pm9.2$	& $1.00$ \\
MACSJ1931.8-2634	& 2.0	& $0.686\pm0.040$	& $1.219^{+0.010}_{-0.019}$	& $3.5\pm0.9$	& $1.00$ \\
RXJ2011.3-5725	& 1.3	& $0.561\pm0.051$	& $0.492^{+0.062}_{-0.107}$	& $2.8\pm0.8$	& $1.00$ \\
MS2053.7-0449	& 0.4	& $0.354\pm0.035$	& $0.513^{+0.127}_{-0.237}$	& $10.2\pm3.1$	& $1.00$ \\
MACSJ2129.4-0741	& -1.4	& $0.401\pm0.024$	& $0.184^{+0.018}_{-0.117}$	& $15.1\pm6.9$	& $1.00$ \\
RXJ2129.6+0005	& 2.0	& $0.544\pm0.028$	& $1.021^{+0.059}_{-0.048}$	& $5.2\pm1.4$	& $1.00$ \\
A2409	& 2.3	& $0.418\pm0.016$	& $0.322^{+0.086}_{-0.233}$	& $9.0\pm2.9$	& $0.65$ \\
MACSJ2228.5+2036	& 0.6	& $0.335\pm0.014$	& $0.182^{+0.008}_{-0.130}$	& $26.2\pm5.2$	& $1.00$ \\
MACSJ2229.7-2755	& 1.3	& $0.686\pm0.082$	& $1.319^{+0.014}_{-0.021}$	& $2.5\pm1.0$	& $1.00$ \\
MACSJ2245.0+2637	& 1.1	& $0.603\pm0.050$	& $0.874^{+0.069}_{-0.052}$	& $3.4\pm0.8$	& $1.00$ \\
RXJ2247+0337	& -0.6	& $0.284\pm0.100$	& $0.020^{+0.245}_{-0.013}$	& $13.3\pm4.2$	& $1.00$ \\
AS1063	& 0.7	& $0.549\pm0.016$	& $0.271^{+0.022}_{-0.017}$	& $7.5\pm1.5$	& $1.00$ \\
CLJ2302.8+0844	& 0.2	& $0.281\pm0.038$	& $0.341^{+0.230}_{-0.316}$	& $25.7\pm9.2$	& $1.00$ \\
A2631	& -0.9	& $0.276\pm0.012$	& $0.066^{+0.088}_{-0.049}$	& $8.9\pm4.6$	& $1.00$ \\

\end{longtable}
\twocolumn
\normalsize

\noindent
population, and is denoted as $\sigma_{L|T}$.

The \LT\ relation was fit for the full cluster population, and the
CC/NCC and relaxed/unrelaxed subsets, and the best fitting parameters
and measured intrinsic dispersion are given in Table \ref{t.ltc}. In
all cases, the normalisation measured for the CC or relaxed clusters
was significantly higher than that of the NCC or unrelaxed
clusters. This is in line with expectations, and the effect is clear
in the distribution of the clusters in the \LT\ plane. The intrinsic
scatter of each of the relaxed/CC subsamples is larger than the
unrelaxed/NCC subsamples, and the highest scatter is that of the
population as a whole. This is consistent with the relaxed/CC clusters
and unrelaxed/NCC clusters forming two fairly distinct populations,
offset in the \LT\ plane.

\begin{figure}
\begin{center}
\scalebox{0.32}{\includegraphics*[angle=270]{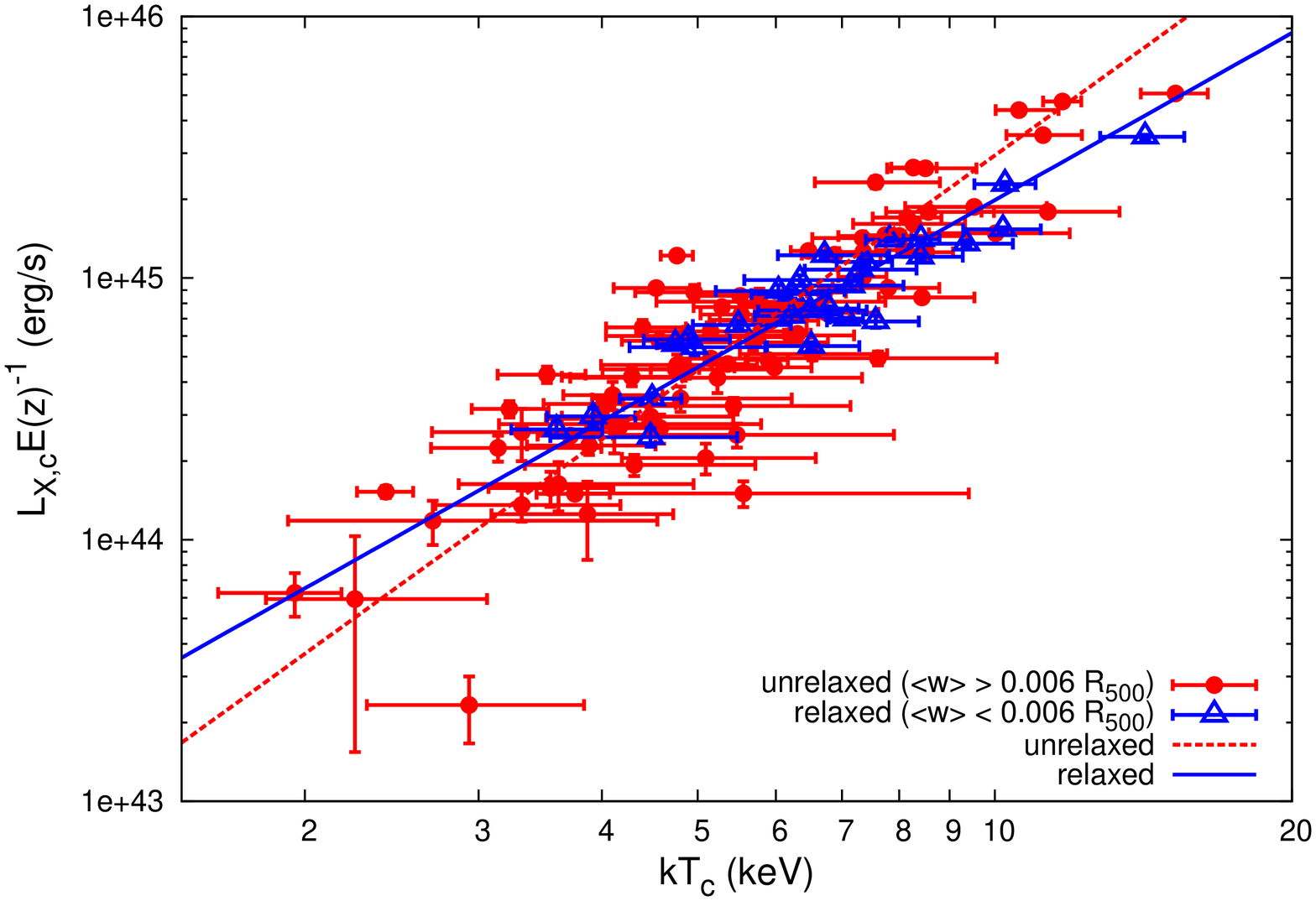}}\\
\scalebox{0.32}{\includegraphics*[angle=270]{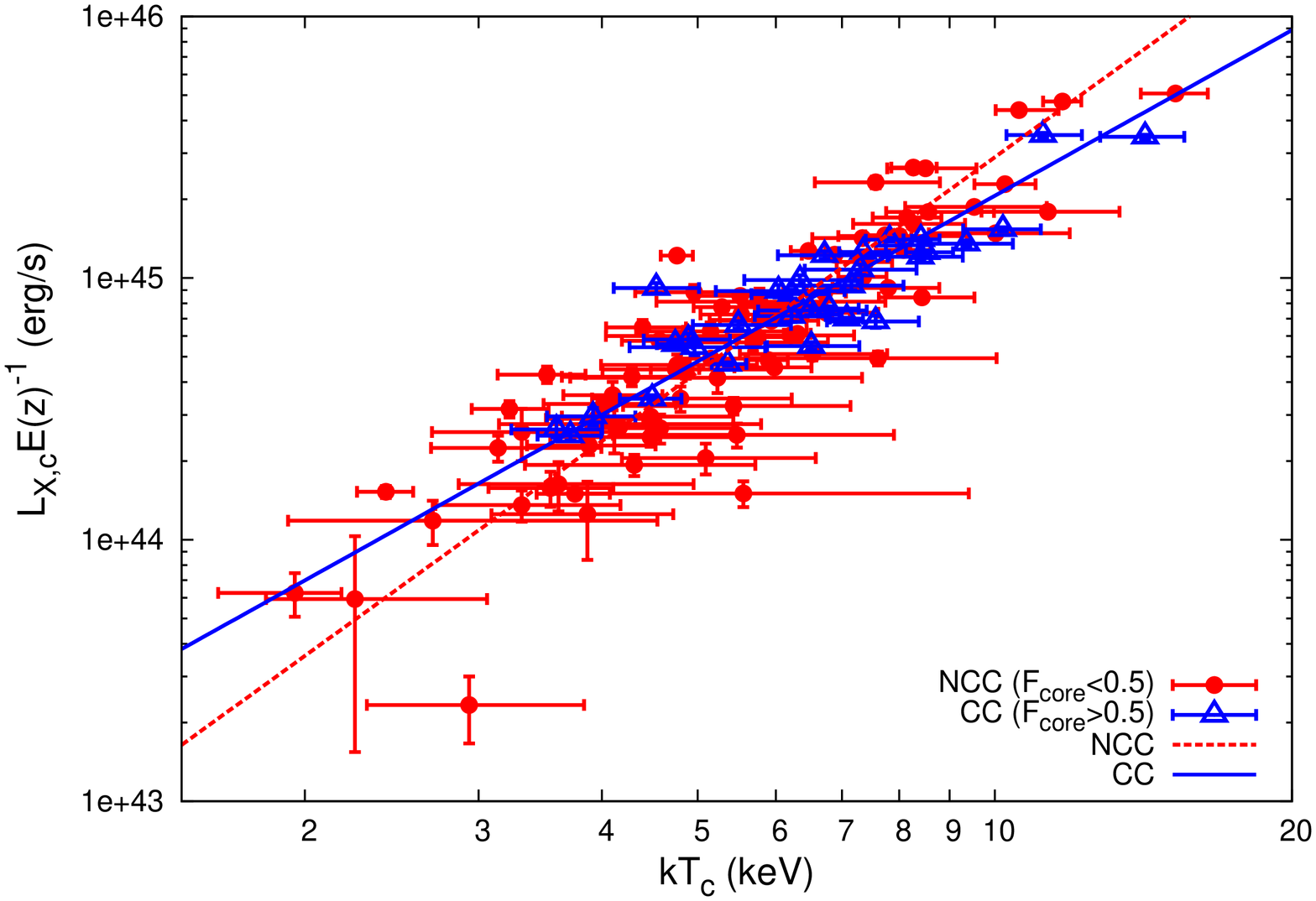}} \\
\caption[]{\label{f.lt} \LT\ relation for the cluster sample, separated
  into relaxed/unrelaxed subsamples (top panel) or CC/NCC subsamples
  (bottom panel). \Lx\ and kT were measured within \rf, with the
  central $0.15\rf$ excluded. Luminosities are corrected for
  self-similar evolution as indicated on the ordinate axis. The lines
  show the best fitting power laws determined from a BCES orthogonal
  regression fit (see text for details).}
\end{center}
\end{figure}

\begin{table*}
\scalebox{1.0}{
\begin{tabular}{llcccc}
\hline
 category        &  filter            &    N  &  A$_{\mathrm{LT}}$ ($10^{44}\ergps$) &  B$_{\mathrm{LT}}$  &  $\sigma_{LT}$ (\%) \\
\hline
all	& none	& 114	& 12.94 $\pm$ 1.00	& 3.63 $\pm$ 0.27	& 67.2 $\pm$ 4.8	\\
CC	& cuspiness $\ge0.8$	& 30	& 26.30 $\pm$ 3.68	& 3.33 $\pm$ 0.60	& 54.6 $\pm$ 4.6	\\
NCC	& cuspiness $<0.8$	& 84	& 9.75 $\pm$ 0.55	& 3.18 $\pm$ 0.20	& 36.8 $\pm$ 4.4	\\
CC	& \Fcore $\ge0.5$	& 31	& 23.28 $\pm$ 2.27	& 2.53 $\pm$ 0.44	& 50.6 $\pm$ 5.6	\\
NCC	& \Fcore $<0.5$	& 83	& 10.02 $\pm$ 0.56	& 3.19 $\pm$ 0.20	& 38.1 $\pm$ 4.4	\\
relaxed	& \cshift $\le0.006$	& 28	& 22.23 $\pm$ 2.98	& 3.10 $\pm$ 0.61	& 61.1 $\pm$ 6.1	\\
urelaxed	& \cshift $>0.006$	& 86	& 10.54 $\pm$ 0.70	& 3.26 $\pm$ 0.22	& 48.5 $\pm$ 6.9	\\
RCC	& combined	& 21	& 27.35 $\pm$ 2.64	& 2.44 $\pm$ 0.43	& 38.0 $\pm$ 3.2	\\
NRCC	& combined	& 93	& 10.57 $\pm$ 0.64	& 3.22 $\pm$ 0.20	& 46.9 $\pm$ 6.0	\\
\hline
\end{tabular}

}
\caption{\label{t.ltc} \LT\ relations for properties measured with core regions
  included ($[0-1]\rf$ aperture). The first column gives the category
  of each subsample, and the second column gives the filter used to
  define that subsample. The combined filter refers to clusters with
  cuspiness$>0.8$, $\Fcore>0.5$, and $\cshift<0.006$. The third column
  gives the number of clusters in each subsample. The fourth, fifth
  and sixth columns give the normalisation and slope of the best
  fitting model and the intrinsic scatter of the data about that
  model. The model used was $\Lx =
  E(z)A_{LT}(kT/T_*)^{B_{LT}}$, with $T_*=6\keV$ in all cases.}
\end{table*}

In order to remove the strong effect of the cool cores, the \LT\
relation was also derived for core-excised cluster properties
(i.e. both \Lx\ and kT measured in the $[0.15-1]\rf$ aperture). The
resulting \LT\ relations are plotted in Figure \ref{f.lt}, again for
the whole population, and for relaxed/unrelaxed and CC/NCC
subsets. The relations were fit as before and the best fitting
parameters are given in Table \ref{t.lt}. In these core-excised
relations, the \LT\ normalisations for all subsamples and the overall
population are in good agreement, indicating that the core excision
has removed the offset between the relaxed/CC and unrelaxed/NCC
populations.

\begin{table*}
\scalebox{1.0}{
\begin{tabular}{llcccc}
\hline
category        &  filter            &    N  &  A$_{\mathrm{LT}}$ ($10^{44}\ergps$) &  B$_{\mathrm{LT}}$  &  $\sigma_{LT}$ (\%) \\
\hline
all	& none	& 114	& 6.98 $\pm$ 0.30	& 2.72 $\pm$ 0.18	& 29.7 $\pm$ 4.2	\\
CC	& cuspiness $\ge0.8$	& 30	& 7.16 $\pm$ 0.33	& 2.15 $\pm$ 0.17	& 13.0 $\pm$ 11.7	\\
NCC	& cuspiness $<0.8$	& 84	& 6.98 $\pm$ 0.39	& 2.82 $\pm$ 0.21	& 33.1 $\pm$ 5.3	\\
CC	& \Fcore $\ge0.5$	& 31	& 7.05 $\pm$ 0.34	& 2.10 $\pm$ 0.13	& 19.2 $\pm$ 5.1	\\
NCC	& \Fcore $<0.5$	& 83	& 7.18 $\pm$ 0.38	& 2.86 $\pm$ 0.22	& 31.4 $\pm$ 6.2	\\
relaxed	& \cshift $\le0.006$	& 28	& 6.71 $\pm$ 0.34	& 2.12 $\pm$ 0.17	& 12.5 $\pm$ 6.5	\\
urelaxed	& \cshift $>0.006$	& 86	& 7.28 $\pm$ 0.39	& 2.86 $\pm$ 0.21	& 32.4 $\pm$ 5.9	\\
RCC	& combined	& 21	& 7.28 $\pm$ 0.26	& 1.90 $\pm$ 0.14	& 1.8 $\pm$ 3.1	\\
NRCC	& combined	& 93	& 7.10 $\pm$ 0.36	& 2.82 $\pm$ 0.19	& 32.5 $\pm$ 5.1	\\
\hline
\end{tabular}

}
\caption{\label{t.lt} \LT\ relations for properties measured with core regions
  excluded ($[0.15-1]\rf$ aperture). The first column gives the category
  of each subsample, and the second column gives the filter used to
  define that subsample. The combined filter refers to clusters with
  cuspiness $>0.8$, $\Fcore>0.5$, and $\cshift<0.006$. The third column
  gives the number of clusters in each subsample. The fourth, fifth
  and sixth columns give the normalisation and slope of the best
  fitting model and the intrinsic scatter of the data about that
  model. The model used was $\Lx =
  E(z)A_{LT}(kT/T_*)^{B_{LT}}$, with $T_*=6\keV$ in all cases.}
\end{table*}

\begin{figure*}
\begin{center}
\scalebox{0.64}{\includegraphics*[angle=270]{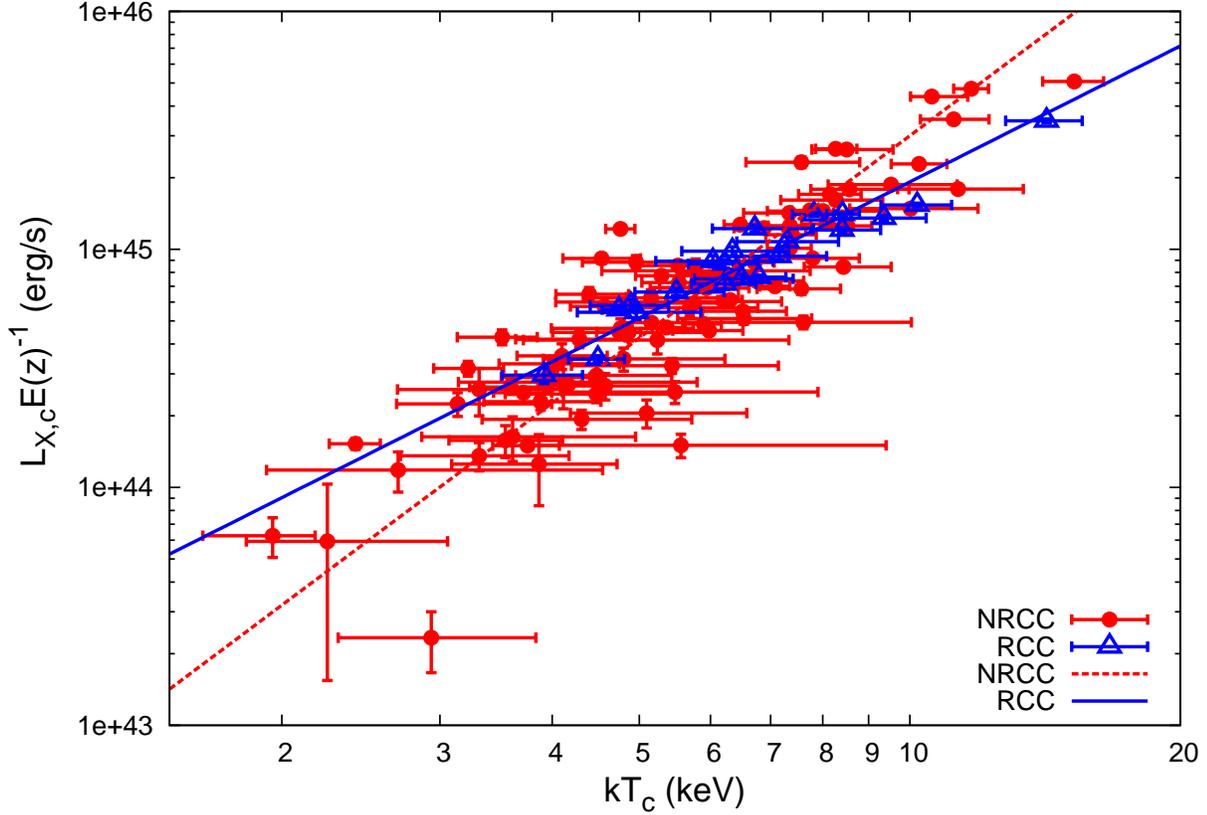}}\\
\caption[]{\label{f.ltcc3} \LT\ relation for the cluster sample, split
  using our combined filter into a subsample containing the most
  relaxed clusters with strongest cool cores (RCC), and the
  complementary subsample (NRCC). \Lx\ and kT were measured within
  \rf, with the central $0.15\rf$ excluded. Luminosities are corrected
  for self-similar evolution as indicated on the ordinate axis. The
  lines show the best fitting power laws determined from a BCES
  orthogonal regression fit (see text for details).}
\end{center}
\end{figure*}

In common with previous studies of luminosity scaling relations
\citep[e.g.][]{mar98a,mau07b,pra09a}, the intrinsic scatter of the
whole population was significantly reduced by the exclusion of the
cores, from $67\%$ to $30\%$ in this case. Indeed, the intrinsic
scatter of each of the subsamples was reduced.  The typical reduction
in dispersion for the unrelaxed/NCC clusters was from $\sim40\%$ to
$\sim30\%$, while for the relaxed/CC clusters the effect was stronger,
with a typical reduction from $\sim50\%$ to $\sim15\%$ (albeit with
larger uncertainties on the intrinsic scatter). Also of note is that
the 21 RCC clusters (corresponding to the most relaxed clusters with
strongest cool cores) have no measureable intrinsic scatter when the
cores are removed, indicating a very regular population. The
core-excluded \LT\ relation for subsamples separated by the combined
filter (RCC and NRCC) is shown in Fig. \ref{f.ltcc3}.

The removal of the core regions also has the effect of reducing the
slope of the \LT\ relation for each subsample, although the effect is
not always significant. The strongest effect is for the population fit
as a whole, and is most likely due to the offset \LT\ relations of the
CC and NCC clusters, combined with different distributions with
temperature of the CC and NCC clusters. Specifically, the CC clusters
which are offset above the rest of the population in the \LT\ plane
are generally hotter in our sample (the median core-excised \kT\ is
$6.5\keV$ for CCs and $5.5\keV$ for NCCs). When the cores are excised,
these clusters move significantly towards lower \Lx\ and higher \kT,
giving rise to the flatter slope than when cores are not excised.

Perhaps of greater interest is that with cores excised, a difference
in \LT\ slope is apparent between the relaxed/CC and unrelaxed/NCC
populations. In all cases, the best fitting slope to the relaxed/CC
clusters is shallower than that of the complementary population (see
Fig. \ref{f.ltcc3}). The significance of the difference ranges from
$\approx2.5\sigma$ to $\approx4\sigma$, with the strongest difference
found for the populations split by the combined filter. It is striking
that all of the relaxed/CC subsamples have \LT\ slopes consistent with
the self-similar prediction of $2$. This is a first indication that
this part of the cluster population obeys self-similar scaling
laws. This was investigated further via an examination of the cluster
gas density profiles.

\subsection{Similarity of gas density profiles}
\begin{figure}
\begin{center}
\scalebox{0.32}{\includegraphics*[angle=270]{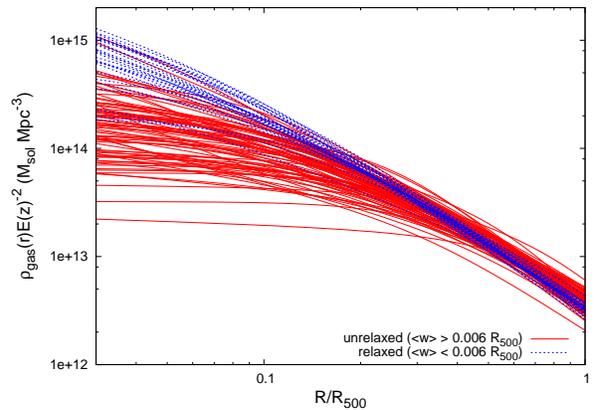}} \\
\caption[]{\label{f.rho} Scaled gas density profiles of the cluster
  sample. The profiles are the best fitting models to the observed
  cluster emissivity profiles (see text for details), and are scaled
  in radius by \rf\ and in density by $E^2(z)$. The solid lines
  indicate the unrelaxed clusters while the dashed lines are the
  profiles of the relaxed clusters.}
\end{center}
\end{figure}

In order to investigate the apparent self-similarity of the \LT\
relation for the relaxed/CC clusters, the structure of the gas density
profiles were examined. The best fitting gas density models for each
cluster were scaled by dividing the radial coordinate by \rf, and
dividing the densities by $E^2(z)$ to remove the expected self-similar
evolution \citep[see e.g.][for examples of the effects of these
scalings]{cro08}. After these scalings, self-similar profiles would be
identical. The scaled profiles are plotted in Figure \ref{f.rho}, with
the relaxed and unrelaxed clusters indicated. In both cases, the
dispersion of the profiles is largest in the central regions, reducing
at larger radii. For both subsamples, the dispersion in the profiles
was significantly larger when the $E^2(z)$ scaling was not applied. It
is apparent that the two subgroups have different gas density
structures, with the relaxed clusters more centrally concentrated, and
the unrelaxed clusters more diffuse.

\begin{figure}
\begin{center}
\scalebox{0.32}{\includegraphics*[angle=270]{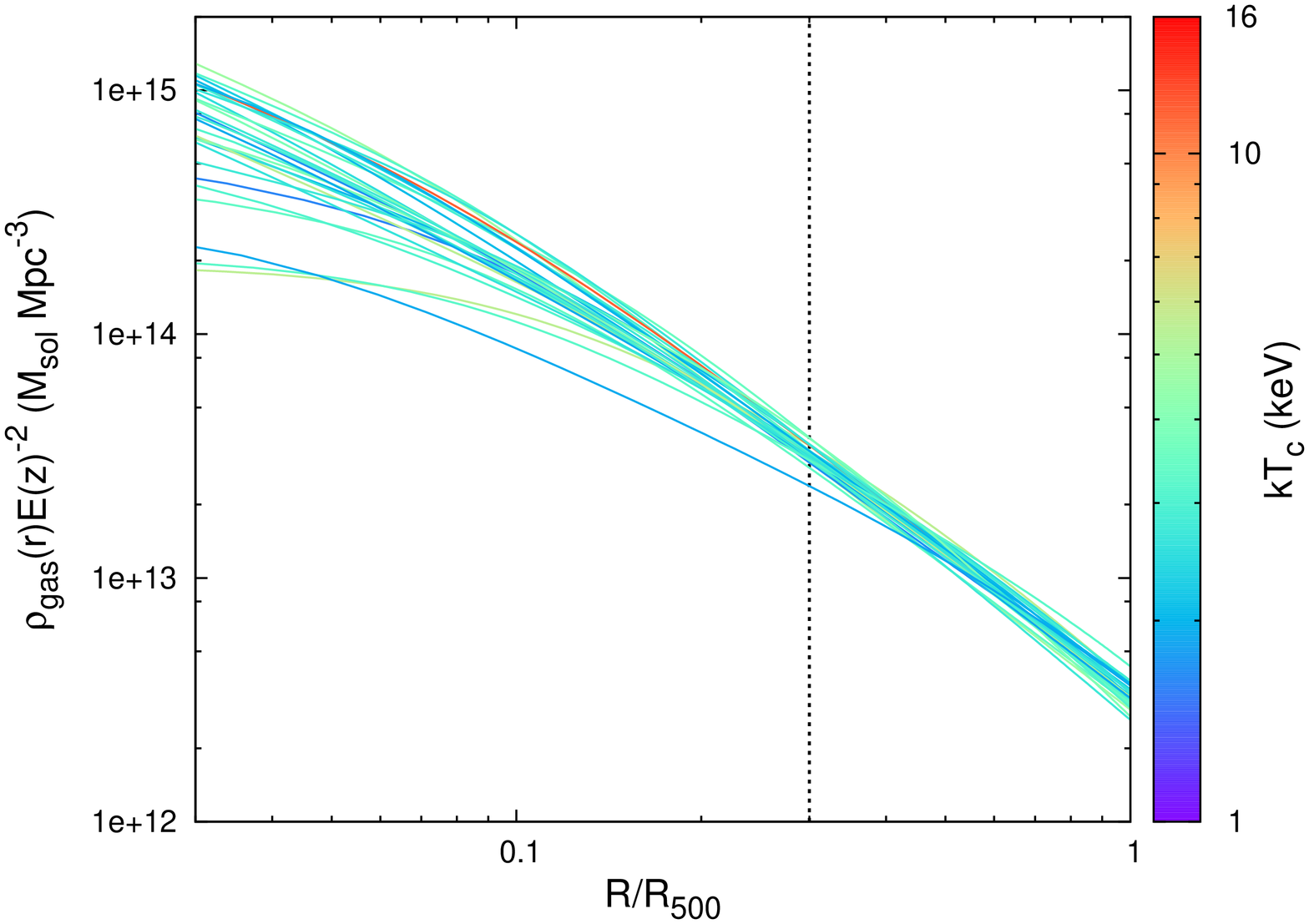}}\\
\scalebox{0.32}{\includegraphics*[angle=270]{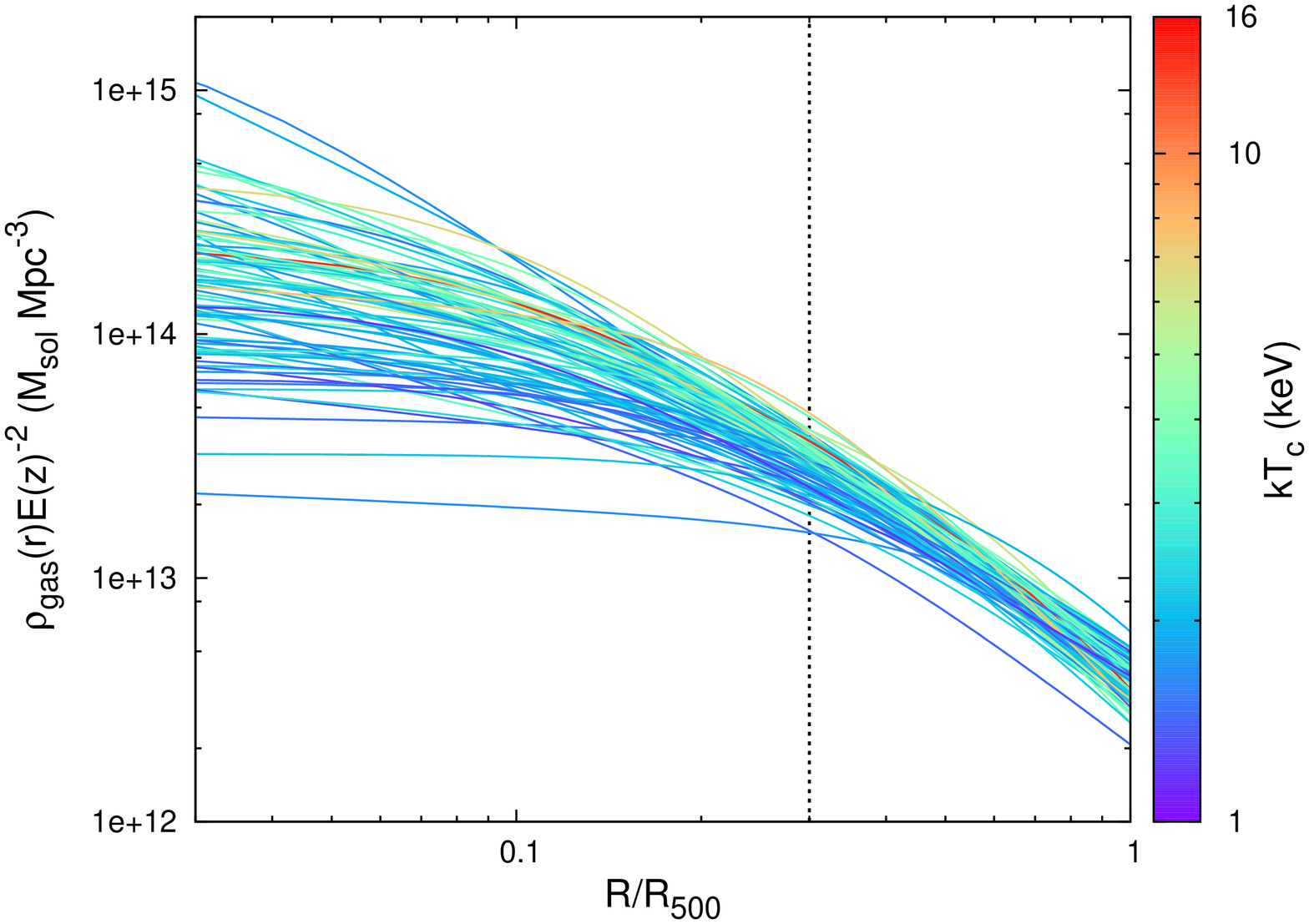}} \\
\caption[]{\label{f.tcol} Scaled gas density profiles of the
  clusters, colour coded by system temperature (as measured in the
  $[0.15-1]\rf$ aperture). The top panel plots the relaxed clusters
  and the bottom panel plots the unrelaxed clusters.}
\end{center}
\end{figure}

\begin{figure}
\begin{center}
\scalebox{0.32}{\includegraphics*[angle=270]{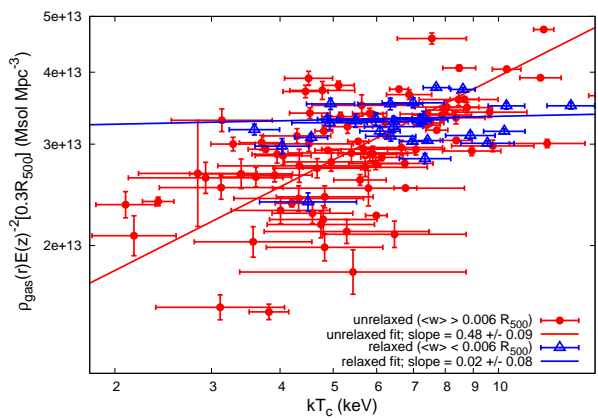}} \\
\caption[]{\label{f.rho-t} Scaled gas density at $0.3\rf$ is plotted
  against system temperature (as measured in the $[0.15-1]\rf$
  aperture). Filled circles and hollow triangles indicate the
  unrelaxed and relaxed clusters respectively.}
\end{center}
\end{figure}

These results demonstrate a lack of similarity between the gas density
profiles of clusters selected as relaxed and unrelaxed. Next, the
degree of self similarity within those subgroups was investigated.
Figure \ref{f.tcol} shows the scaled gas density profiles of the
relaxed and unrelaxed clusters, colour-coded by system temperature
(measured with the core excised). For the unrelaxed clusters, there is
evidence for a temperature dependence of the profiles, with cooler
systems tending to have lower densities than hotter systems at a given
radius. For the relaxed clusters, no such trend is apparent. The
temperature dependence of the density profiles was examined in more
detail by measuring the scaled density of each cluster at a fixed
fraction of \rf, and plotting them against system temperature. An
example plot for densities measured at $0.3\rf$ is shown in Figure
\ref{f.rho-t}. Despite the different temperature distributions of the
two populations, the unrelaxed clusters clearly have a stronger
dependence of gas density on temperature than the relaxed clusters.

\begin{figure}
\begin{center}
\scalebox{0.32}{\includegraphics*[angle=270]{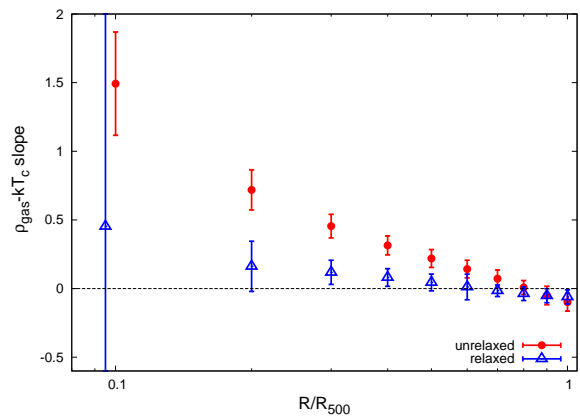}} \\
\caption[]{\label{f.simprof} Similarity index (the slope of the power
  law dependence of scaled gas density on system temperature) is
  plotted as a function of scaled radius. Filled circles and hollow
  triangles indicate the unrelaxed and relaxed clusters
  respectively. The innermost points are slightly offset in radius for
clarity.}
\end{center}
\end{figure}

The data for each subsample were then fit with a power law of the
form:
\begin{eqnarray}
  \label{eq.rho-t}
  \rho(r/\rf)E^{-2}(z) & = & (kT)^\eta,
\end{eqnarray}
where we refer to $\eta$ as the ``similarity index'' of the density
profiles ($\eta=0$ corresponds to self-similar profiles). This
procedure was repeated at different values of $r$ to produce the
similarity index profile shown in Figure \ref{f.simprof}. The
similarity index profile shows that the gas density profiles of the
unrelaxed clusters have a strong temperature dependence in the core
regions that becomes weaker with radius, approaching self-similarity
by $\sim0.7\rf$. A similar trend of decreasing similarity index with
radius was observed by \citet{cro08} who found
values of $0.5$ and $0.25$ at $0.3\rf$ and $0.7\rf$
respectively for their representative sample of clusters.

In contrast, the similarity index profile of the relaxed clusters is
quite flat, with no strong dependence on temperature at any radius. As
a test of the sensitivity of this difference in similarity index to
the absence of relaxed clusters with low temperatures, unrelaxed
clusters with kT$<3.5\keV$ were excluded from the fits. This made no
significant difference to the slopes.

Very similar results were found for density profiles when the CC and
NCC samples were compared, although we consider the separation on
\cshift\ to be preferable as this is less directly linked to the
radial structure of the gas density than either the \Fcore\ or
cuspiness measurements.

\section{Evolution of the \LT\ relation}
Thus far, it has been assumed that the redshift evolution of the \LT\
relation is self-similar, \ie the slope is independent of redshift,
while the normalisation increases with increasing redshift due to the
increasing density of the clusters. As discussed in the introduction,
there is some debate in the literature as to whether this evolution model
is a good description of observed clusters, with evidence that
selection biases play a significant role in evolution studies.

To investigate this further, the evolution in the \LT\ relation of the
current \Chandra\ cluster sample was examined. In order to measure
evolution, a low redshift baseline must be defined. While many exist
in the literature, the limited fields of view of \Chandra\ and \XMM\
mean that at $z<0.1$, the available \LT\ relations were derived from
earlier missions such as \ROSAT\ and \ASCA. For internal consistency,
the \LT\ relation of the clusters in the sample with $0.1<z<0.2$ was
fit as described above and used as the local baseline for our
evolution study. In this fit, and all of the following evolution
study, the core-excised properties were used. The parameters for the
local \LT\ relation were $A_{LT}=(6.05\pm0.04)\times10^{44}\ergps$ and
$B_{LT}=2.51\pm0.29$.

\begin{figure}
\begin{center}
\scalebox{0.32}{\includegraphics*[angle=270]{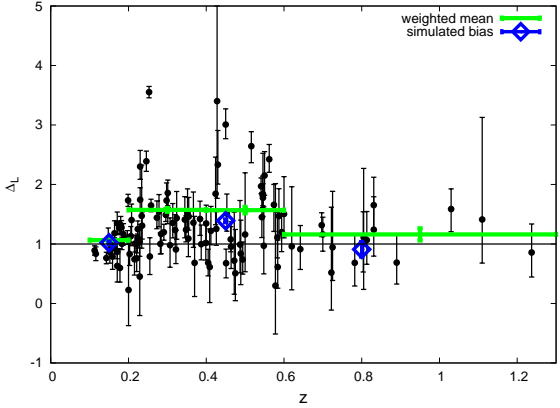}} \\
\caption[]{\label{f.dl} The ratio of the observed, core excised
  luminosity to that predicted by our local \LT\ relation, including
  self similar evolution, is plotted as a function of
  redshift. Clusters evolving self-similarly should scatter around
  $\Delta_L=1$. Also shown are the weighted mean $\Delta_L$ in several
redshift bins, and the possible effect on $\Delta_L$ due to Eddington
bias in our simplistic bias simulations.}
\end{center}
\end{figure}

For each cluster, the ratio of its observed luminosity to that
predicted by self similar evolution of the local relation was
computed, and termed $\Delta_L$. Figure \ref{f.dl} shows the plot of
$\Delta_L$ against redshift. In this plot, clusters following
self-similar evolution would have $\Delta_L=1$. Clusters more (less)
luminous than predicted by the self-similar evolution of our local
relation have $\Delta_L>1$ ($\Delta_L<1$). Figure \ref{f.dl} shows
that relative to the local subset of clusters, those at intermediate
redshifts ($0.2<z<0.6$) are significantly more luminous on average
than the self-similar evolution model predicts. At high redshifts
($z>0.6$) the clusters agree with the self-similar evolution model.

Similar results were reported by \citet[][based in part on data from
M08]{bra07}, and was suggested as evidence against the self-similar
model. However, as noted by \citet[][]{bra07}, selection bias could be
responsible for distorting the observed evolution \citep[see also
\textsection 5.2 in ][]{mau07b}.

At this point, we note that there is some ambiguity in the terms used
to describe different selection biases in the literature on cluster
scaling relations, and so we pause for a brief discussion of
selection bias and terminology.

\subsection{Selection biases and terminology}
We consider two types of bias that affect flux-selected samples of
objects. Either or both of these biases are often referred to simply
as ``Malmquist bias'' in the literature, while occasionally the term
``Eddington bias'' is also used. \citet{tee97} gives a detailed
overview of these selection biases and advocates the use of the terms
Malmquist bias of the first and second kind. For clarity, we use the
term ``Malmquist bias'' to refer to the effect that arises due to more
luminous objects being detectable to greater distances, and
``Eddington bias'' to refer to the effect due to the differential
scattering of objects across the threshold of a flux-limited
sample. We will now discuss each of these biases in more detail in the
following.

Let us consider a population of objects whose luminosities $L$ are
correlated to their masses $M$, with some intrinsic scatter in
luminosity at a given mass (parametrised by $\sigma_{L|M}$). A sample
of these objects whose measured flux is greater than some flux limit
$F_{lim}$ is then defined. In the idealised case where there is no
scatter, the Malmquist bias will lead to the most massive, luminous
objects being over-represented in the sample, due to the larger search
volume in which they may be detected. This is easily corrected for
with knowledge of the survey selection function [the survey volume as
a function of luminosity; $V(L)$] to weight the objects by their
search volume. In the more realistic case when there is scatter in $L$
for fixed $M$, those objects with higher than average $L$ for their
mass, will be more numerous in the sample than those with lower than
average $L$, again due to the larger survey volume. The mean
luminosity of objects in the sample is thus biased higher than the
population average and the size of the bias
depends on $\sigma_{L|M}$.

Turning now to the Eddington bias, let us consider objects close to
the flux limit (by ``close'' we mean objects whose scatter in $L$ has a
non-negligible chance of moving them across the flux limit). Near the
flux limit there exist objects whose mass corresponds to an average
$L$ that would be below the flux limit at their redshift, but whose
observed $L$ is higher than average, and so are included in the sample
(we refer to these as overluminous). Similarly there exist objects
whose mass corresponds to an average $L$ that would be above the flux
limit at their redshift, but whose observed $L$ is lower than average,
and so are not included in the sample (we refer to these as
underluminous). This means that objects included in the sample are
biased to above average $L$. So far, this is equivalent to a special
case of the Malmquist bias, with the survey volume dropping to zero
for objects whose observed luminosity is below the flux limit. What
distinguishes the Eddington bias is an additional factor that
arises when the number density of objects is a decreasing function of
$M$ and $L$ (as is the case for galaxy cluster mass and luminosity
functions). In this case, there are larger numbers of lower mass,
overluminous objects which are included in the sample than higher
mass, underluminous objects which drop out of the sample. This gives a
further bias towards above-average luminosity in the final sample. The
size of the Eddington bias depends on $\sigma_{L|M}$, and on the slope
of the mass function at the mass whose average luminosity corresponds
to the flux limit at the redshift in question.

In the above discussion, we assumed that $\sigma_{L|M}$ was due solely
to an intrinsic dispersion in $L$ at fixed $M$. Similar biases occur if
$L$ is subject to measurement errors. In this case, even in the
absence of intrinsic variation of $L$ at fixed $M$, there is an
observed variation in $L$ which has the same effect as the biases
described above. A complicating factor is that the size of
$\sigma_{L|M}$ will typically depend on $L$ due to the larger
measurement errors on fainter systems. This increases the difference
between the number of overluminous objects moving into the sample from
below the flux limit and underluminous objects moving out of the
sample, further increasing the bias. If the objects are reobserved the
average $L$ will be unbiased for fixed $M$, as the original $L$ values
were randomly drawn from the possible $L$ distribution (this is in
contrast to the case where the variation in $L$ at fixed $M$ is
intrinsic). The number of objects in the sample will remain biased,
however, unless objects below the flux limit are also reobserved, and
the sample redefined.

\subsection{The effect of selection bias on cluster scaling relations}
In studies of the evolving mass function of clusters based on X-ray
flux limited samples, both the Malmquist bias and Eddington bias must
be accounted for. Typically, the Malmquist bias as defined here is
included in modelling the survey selection function, while the
Eddington bias is taken into account when modelling the mass function
\citep[see e.g. ][for useful discussions of the effects of these
biases]{mar98a,vik09,man10}. When X-ray selected clusters of galaxies
are used for studying the scaling relation between \Lx\ and another
property (such as kT), both the Malmquist and Eddington biases will
enhance the mean \Lx\ for a given kT (or mass). The effect of this can
be to bias the normalisation of the \LT\ relation high, and bias the
slope low. The effect on the slope arises because the lower
temperature systems are closer to the flux limit at a given redshift,
and so will suffer stronger bias. This effect is lessened if a wide
redshift range is considered, as higher temperature objects will also
tend to be close to the flux limit. As described above, the size of
the total bias in the \LT\ relation will depend on $\sigma_{L|M}$, and
on the slope of the mass function at the flux limit of the
survey. Note that while $\sigma_{L|M}$ can be significantly reduced by
excluding cluster cores for analysis purposes, the core-included \Lx\
must be used in consideration of bias, as generally clusters are
selected (or not) on the basis of their total \Lx.

In the sample presented here, the clusters are drawn from a
heterogeneous archive, and do not represent a statistically complete
sample. However, most of the clusters were originally detected in
X-ray surveys, and are thus subject to Malmquist and Eddington bias. A
rigorous treatment of the bias is impossible without a well-defined
selection function, but some simple simulations were performed to
investigate the plausible effect that selection bias could have on
the observed evolution of the \LT\ relation in this sample. The
process used was to simulate a self-similar population of clusters,
and apply selection functions representative of different cluster
surveys. As discussed in \citet{mau07b}, we approximate the selection
of clusters in our sample by three very broad classes of X-ray
surveys: at low redshift ($z\approx0.15$) many clusters were selected
in shallow all-sky X-ray surveys, for which we use the Brightest
Cluster Survey \citep[BCS; ][]{ebe98} flux limit as a template (a
bolometric flux of $8.3\times10^{-12}\flux$); at moderate redshift
($z\approx0.45$) most of the clusters come from deeper surveys for
which we use the MACS \citep[][; bolometric flux limit of
$3.4\times10^{-12}\flux$]{ebe01b} as a template (our sample contains
24 MACS clusters with a median redshift of $0.42$); at high redshifts
($z\approx0.85$) most clusters in our sample were detected in deep
surveys of \ROSAT\ pointings, for which we use the WARPS
\citep[][bolometric flux limit of $9.1\times10^{-14}\flux$]{sch97} as
a template.

\begin{figure}
\begin{center}
\scalebox{0.85}{\includegraphics*[angle=0]{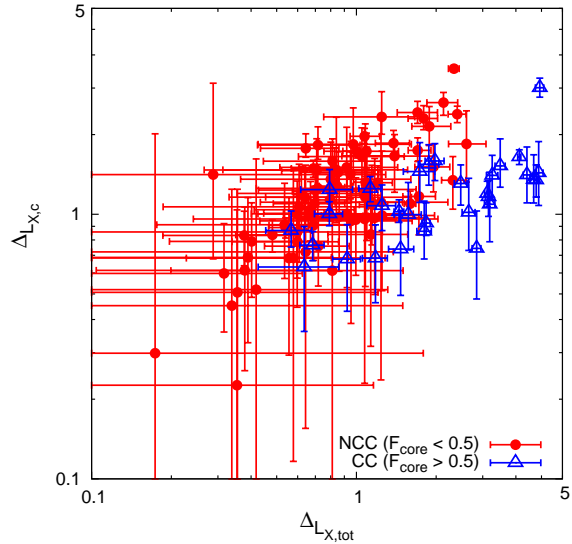}}
\caption[]{\label{f.dev-devc} Plot showing the correlation between the
  ratio of observed to predicted \Lx, when \Lx\ and kT are measured
  with ($\Delta_{\Lxt}$) and without ($\Delta_{\Lxc}$) the core
  regions included.}
\end{center}
\end{figure}

At each of the approximate redshifts corresponding to the above survey
classes, a mass function was constructed using the model of
\citet[][see their Equation 9]{jen01}.  This model is based on fitting
results from cosmological simulations, and is similar in form to the
standard Press-Schechter mass function \citep{pre74}, although it
provides a significantly improved fit to the simulations.  The
redshift range for each survey class was broken into smaller redshift
bins, and clusters were drawn with random redshifts and masses (from
the above mass function) until we obtained a set number of detected
clusters within each bin (where a cluster is ``detected'' if its
scattered total flux falls above the flux limit for the appropriate
survey).  The cluster luminosities were determined as follows. Each
cluster was assigned a total (i.e. core included) luminosity \Lxt\ and
a core excised luminosity \Lxc, using the $\LM$ relations of
\citet{mau07b}. Importantly, these relations include self similar
evolution, so the luminosity at fixed temperature evolves smoothly as
$E(z)$. \Lxt\ was then randomised under a lognormal distribution with
$\sigma=39\%$ to give $\Lxt'$, reproducing the intrinsic dispersion of
the \citet{mau07b} $\Lxt-M$ relation. The value of \Lxc\ was also
shifted in the same direction by the same fractional amount but scaled
by $17/39$ \citep[the ratio of the intrinsic scatters in the][\Lxc-M
  and \Lxt-M relations]{mau07b} to give $\Lxc'$. This gives consistent
values of the total and core excised luminosities for each cluster,
including intrinsic scatter. By scaling \Lxc\ in this way, we are
making the assumption that the dispersion in \Lxt\ is correlated with
that in \Lxc. This is supported by Fig. \ref{f.dev-devc}, which shows
$\Delta_{\Lxt}$ plotted against $\Delta_{\Lxc}$ for the observed
clusters (where $\Delta_{\Lxt}$ is the ratio of the observed \Lxt\ to
that predicted by the low-redshift $\Lxt-$kT relation, and
$\Delta_{\Lxc}$ is the same but for core-excised quantities), and shows
a positive correlation. The weak correlation for the CC clusters in
the plot is testament to the efficacy of removing the core of those
systems in reducing the dispersion in the \LT\ relation.
Finally, for each cluster, a temperature was assigned,
derived from the \citet{vik06a} \MT\ relation, again assuming self
similar evolution, but neglecting any intrinsic scatter. Thus, at each
of our redshifts of interest, we have produced a population of
clusters whose observed properties ($\Lxt'$, $\Lxc'$ and kT) are
(by construction) described perfectly by a consistent set of self-similarly evolving
scaling relations, with intrinsic dispersion in \Lx.

The flux limit of each of our three representative surveys was
converted to a luminosity limit at the appropriate redshift, and
clusters with $\Lxt'$ below that limit were rejected. In this way, the
effects of the selection biases are reproduced. For the clusters
surviving the survey selections, the ratio of $\Lxc'$ to $\Lxc$ was
calculated, which corresponds to the $\Delta_L$ of the observed
clusters. Figure \ref{f.dl} shows the mean $\Delta_L$ for the
simulated clusters at each redshift. The departures from self similar
evolution ($\Delta_L=1$) are solely due to the selection biases, and as
the intrinsic scatter $\sigma_{L|M}$ was constant with redshift in our
simulations, the variation in $\Delta_L$ with redshift is simply due
to the shape of the mass function close to the flux limit of the
different surveys and redshifts considered here.

\section{Discussion}
\subsection{Strong self-similarity}
The examination of the slopes of the \LT\ relation in different
sub-populations of this large cluster sample indicated that the most
relaxed systems, or those hosting the strongest cool cores
(definitions with significant overlap), exhibit a low scatter \LT\
relation with a slope of $\approx2$, in agreement with the
self-similar model, once their core regions have been excised. Such a
self-similar correlation has generally not been found in previous
studies of the \LT\ relation, which is likely due our large sample
size, and to the fact that most previous studies have not
distinguished CC and NCC subsamples once a core correction has been
made. However, \citet[][hereafter P09]{pra09a} did make such a
distinction in their recent \XMM\ study of the \LT\ relation of the
REXCESS sample, and used a similar methodology to us allowing
straightforward comparisons to be made.

In the P09 study, after core exclusion, the \LT\ slopes of the
relaxed/CC subsamples were shallower than the unrelaxed/NCC
subsamples, but (in contrast to our results) were still significantly
steeper than the self-similar slope of 2. The methods used to define
the CC/NCC and relaxed/unrelaxed classes in P09 differ slightly from
those employed here. P09 separated CC and NCC clusters using a
measurement of the central density, and classed $\sim30\%$ of their
clusters as CC (a similar fraction to us). To classify relaxed and
unrelaxed clusters, P09 used the same centroid shift measurement as
us (albeit with a slightly different implementation), but used a value
of $\cshift=0.01\rf$ to segregate the relaxed/unrelaxed subsets. This
less strict definition of a relaxed cluster than our
$\cshift>0.006\rf$, resulted in a significantly larger fraction of the
P09 sample ($\sim60\%$) being classed as relaxed.

\begin{figure}
\begin{center}
\scalebox{0.32}{\includegraphics*[angle=270]{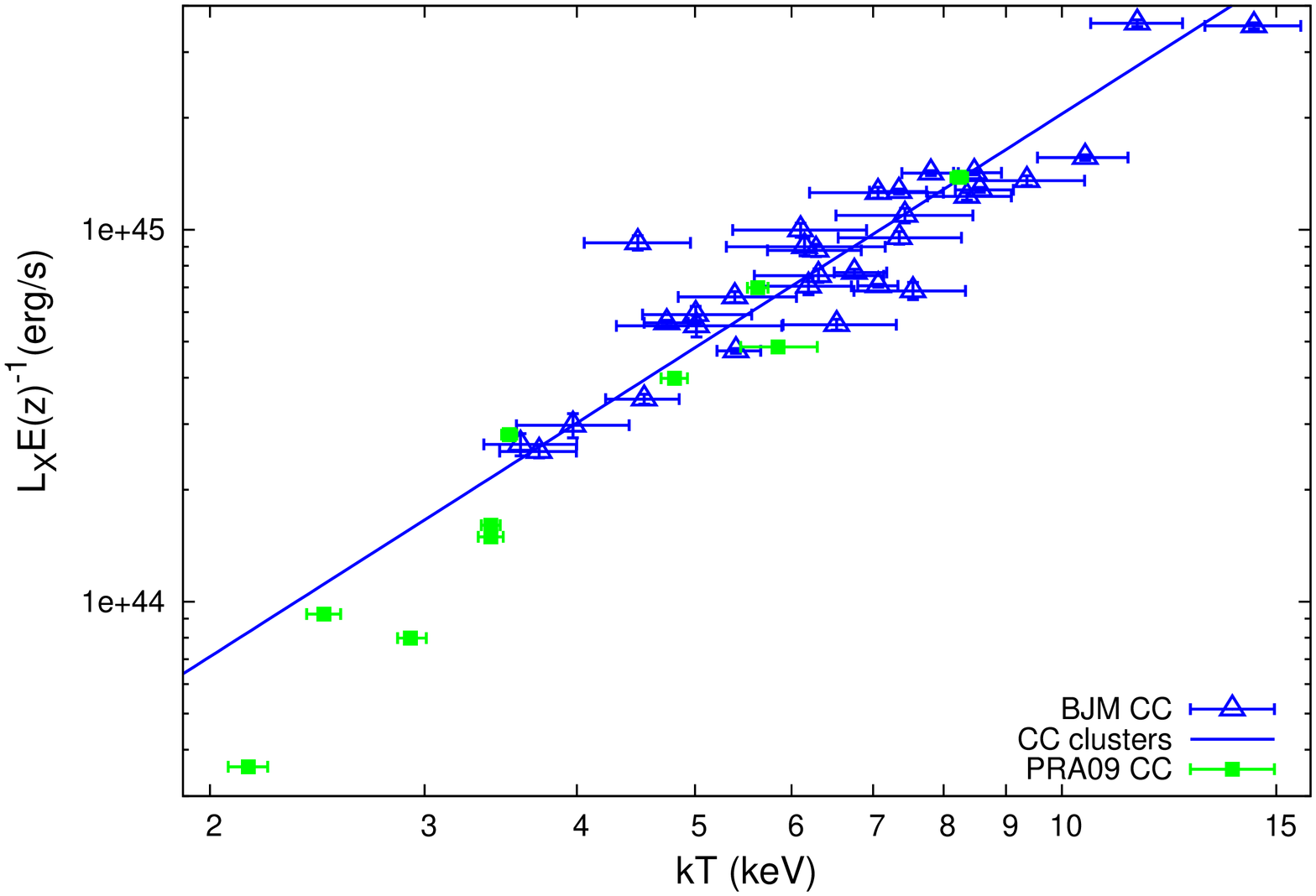}}\\
\scalebox{0.32}{\includegraphics*[angle=270]{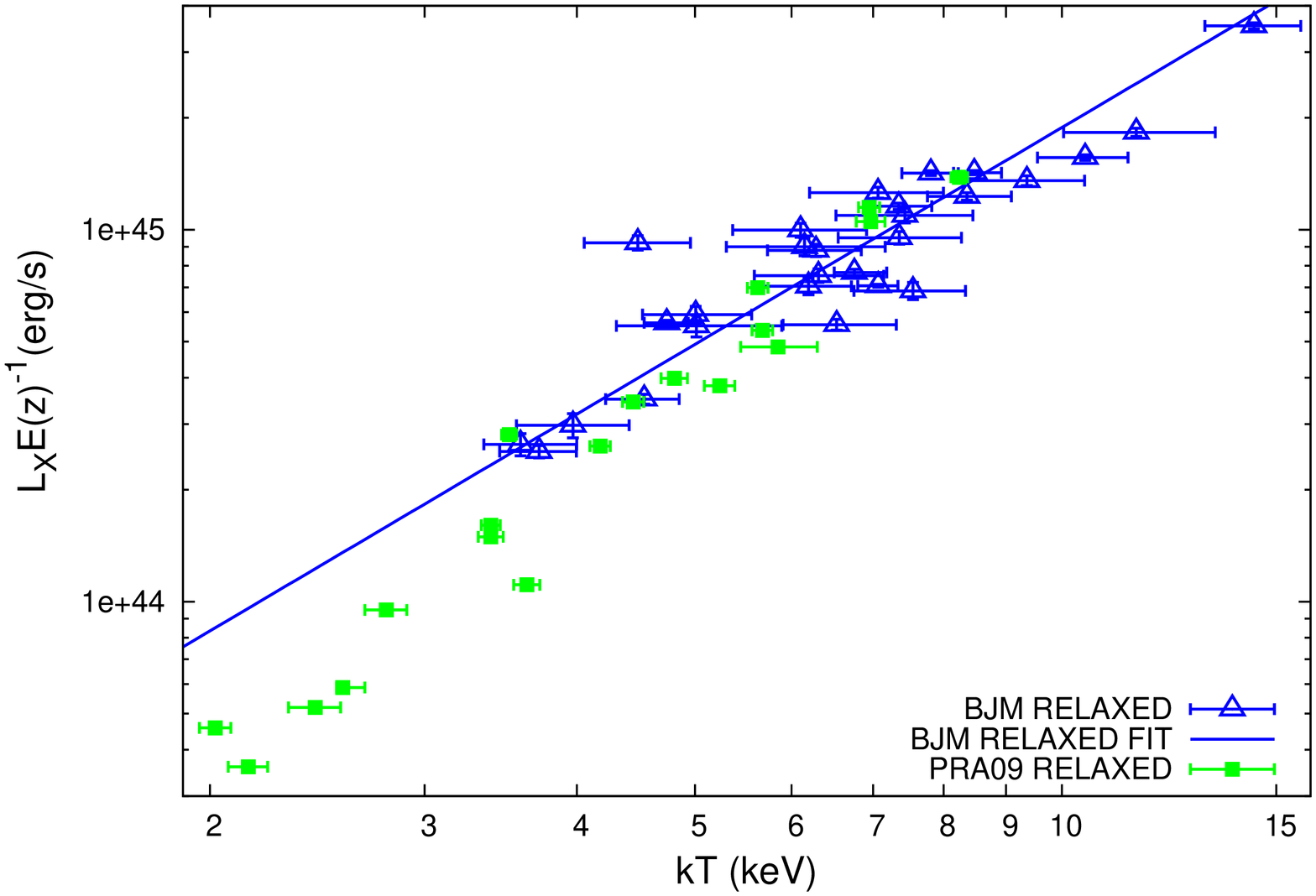}}
\caption[]{\label{f.pra09} {\it Top:} the core-excised \LT\ relation for the CC clusters in
  the current sample is plotted along with the CC clusters of
  P09. {\it Bottom:} the core-excised \LT\ relation for the relaxed clusters in
  the current sample is plotted along with the relaxed clusters of P09.}
\end{center}
\end{figure}

Subsample definition and instrumentation notwithstanding, the key
difference between the P09 and our samples is the temperature range of
systems covered. Half of the P09 CC clusters are cooler than $3.5\keV$
(our coolest CC system). This is made clear in Figure \ref{f.pra09},
which plots the relaxed and CC P09 clusters on the corresponding \LT\
relations of our sample. At $\kT\ga3.5\keV$ the P09 data agree quite
well with our \LT\ relations, with their essentially self-similar
slopes (particularly for the CC selection). This suggests that the
steeper P09 \LT\ slope is driven by their inclusion of lower mass
systems, and that the self-similarity of relaxed/CC clusters found
here is limited to more massive systems ($\kT>3.5\keV$).

\begin{figure}
\begin{center}
\scalebox{0.32}{\includegraphics*[angle=270]{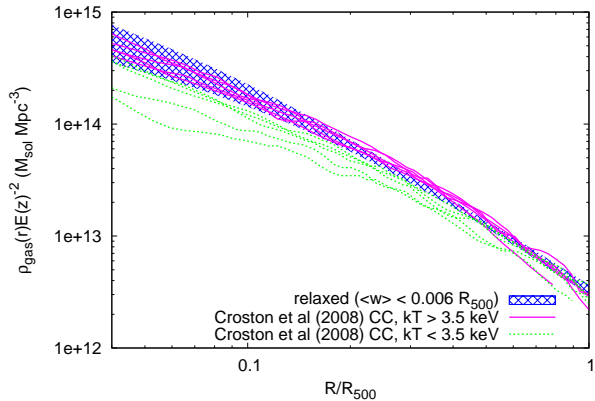}}
\caption[]{\label{f.cro08prof} Scaled gas density profiles of the
  REXCESS CC clusters \citep[taken from][]{cro08}. The REXCESS
  clusters are divided into hot and cool subsets as indicated. The
  hatched region indicated the standard deviation of the scaled gas
  density profiles of the relaxed clusters in our sample.}
\end{center}
\end{figure}

\citet{cro08} presented the gas density profiles of the REXCESS
clusters, allowing us to investigate if this possible similarity
breaking of the \LT\ relation below $3.5\keV$ is manifested in the gas
density profiles. In Fig. \ref{f.cro08prof} we show the scaled gas
density profiles of the CC REXCESS clusters, separated into hot
($\kT>3.5\keV$) and cool ($\kT<3.5\keV$) subsets. Also indicated is
the mean profile of the relaxed clusters from our sample (as plotted
in Fig \ref{f.rho}). The hot REXCESS profiles agree fairly well with
our mean relaxed profile (which are also all $\kT>3.5\keV$ clusters),
but the cool REXCESS CC clusters have flatter profiles, with lower gas
densities than the other cluster profiles out to $\sim0.7\rf$. This
suggests that the self-similarity observed in the gas density profiles
of the relaxed clusters in our sample is valid only for hotter
($\kT>3.5\keV$) systems. This is consistent with the proposed
steepening of the \LT\ relation below $3.5\keV$.

The scaled gas density profiles of our large sample also provide
information on where self-similarity breaks down. For the
unrelaxed/NCC clusters, these profiles showed that the similarity
index was $>0$ out to a cluster-centric radius of
$\approx0.7\rf$. Beyond this radius, the gas density profiles of the
unrelaxed/NCC clusters appear self-similar. These results broadly
agree with the \XMM\ measurements of \citet[][although in that work,
clusters were not separated into relaxed/unrelaxed
subsamples]{cro08}. In contrast, the similarity index of the
relaxed/CC clusters is consistent with zero at all radii probed
(Fig. \ref{f.simprof}). This would seem to imply that excluding the
cores should not be required to recover a self-similar \LT\
relation. This may be the case, but as illustrated in Figures
\ref{f.rho-t}, while there is no strong temperature dependence of the
scaled gas density in the cores, there is departure from
self-similarity in the form of large cluster-to-cluster variation in
scaled gas density. This is indicative of different cool core
strengths and is likely responsible for the non self-similarity of the
core-included \LT\ relation. Finally, the REXCESS CC profiles in
Fig. \ref{f.cro08prof} show that in cool ($\kT<3.5\keV$) CC systems,
the core gas density is lower than in hotter systems, suggesting that
a non-zero similarity index would be measured at radii smaller than
$\approx0.7\rf$ for the relaxed/CC clusters in our sample if it
included cooler systems.

Outside a radius of $\sim0.15\rf$, the similarity index of relaxed/CC
clusters is close to zero and the cluster-to-cluster variation in
scaled gas density is low, indicating that the ICM in these regions is
self-similar. This is consistent with the self-similar slope of the
core-excised \LT\ relation of the relaxed/CC clusters.

Our results can also be usefully compared with the study by
\citet{man10a} who found that when core regions were excised, the
\LT\ relation slope was $2.70\pm0.20$.  \citet{man10a} did not
separate relaxed and disturbed or CC/NCC clusters in their analysis,
but their slope is in excellent agreement with
the slope we find for all clusters with cores excluded
($2.72\pm0.18$). Usefully, \citet{man10a} were able to include
full treatment of selection biases in their analysis, and
showed that for core-excised luminosities, selection biases did not
significantly impact the measured scaling relations, so the
self-similar slope we find for the core-excised \LT\ relation of the
relaxed/CC clusters in our sample is not likely to be affected by
selection bias.

We thus conclude that the ICM of massive (kT$>3.5\keV$), relaxed (or
CC) clusters obeys strong self-similarity outside the core
($r<0.15\rf$) regions.

\subsection{Implications for similarity-breaking models}
\citet{pra10} argued that variations of the gas content of
clusters with mass and radius are at the root of the observed
departures from self-similarity of cluster entropy profiles. They
proposed a mechanism whereby gentle heating from AGN, coupled with
merger related mixing is responsible for a redistribution of gas,
leading to the observed entropy profiles, and also the suppression of
luminosity in low mass systems, giving rise to steepening in the \LT\
relation. Our results are consistent with this picture, but add the
requirement that the most massive relaxed/CC systems are essentially
self similar outside the central $0.15\rf$. The entropy of the CC
clusters in the \citet{pra10} sample shows evidence for excess above
the self-similar expectations at radii as large as $\sim0.7\rf$,
however, this is driven by the lower mass systems. Clusters with
kT$>3.5\keV$ are in reasonable agreement with the self-similar entropy
model beyond $\sim0.15\rf$ \citep[roughly $0.1\rt$, as plotted in
figure 3 in][]{pra10}.

Our results on the details of the \LT\ relation and density structure
of the ICM present new information with which to refine models of ICM
heating and feedback. In particular, by identifying a fully
self-similar regime in the cluster parameter space, we have a baseline
against which to measure the impact of feedback.

Let us first consider the case of relaxed/CC clusters. With the core
regions excluded, we find a self-similar \LT\ relation, which appears
(when combined with the P09 data) to steepen below $\sim3.5\keV$. The
density profiles of the $\kT>3.5\keV$ relaxed/CC systems are also
self-similar beyond the core, with low dispersion and similarity index
close to zero, but again the cooler REXCESS systems show similarity
breaking, with suppressed densities relative to the hotter
systems. Our results thus suggest that for relaxed systems above
$\sim3.5\keV$, the effects of feedback are negligible outside the core
of $0.15\rf$.

A possible interpretation of this is that feedback occurs in the cores
of all of the relaxed/CC systems, providing some fixed level of energy
input beyond the balancing of the gas cooling. The heated gas will
have raised entropy, and will rise, expand and cool, conserving
entropy, until it reaches a radius of gas with the same entropy
level. For more massive systems, this will occur at a smaller radius
\citep[see e.g. figure 1 in][]{pra10}. Thus for a mass-independent
amount of heating, there will be a mass threshold above which the
effects of feedback are confined to the core regions. For a $3.5\keV$
system, the entropy level at $0.15\rf$ is $\sim200\keV\cc$
\citep{pra10}, so an entropy increase somewhat lower than this would
not redistribute gas much beyond the core. In the absence of mergers,
the new entropy structure will remain intact until ``eroded'' by
radiative cooling, but periodic core feedback would maintain a
quasi-static ICM configuration.

For unrelaxed/NCC clusters, the population does not appear to be
self-similar at any mass, even with the cores excluded. This is
manifested in the steep \LT\ relation which crosses that of the
relaxed/CC clusters at $\sim6\keV$, and also in the gas density
profiles, which show temperature dependence out to large scaled
radii. The steepening of the \LT\ relation below $6\keV$ can be
explained in the same terms as for the $<3.5\keV$ relaxed/CC clusters
if we posit that core feedback occurs in all clusters (not just
relaxed/CC) clusters. This is consistent with quasar mode heat input
occurring in all clusters \citep{sho09}. We then require that the
effects of feedback be noticeable outside the core for more massive
systems (up to $\kT\approx6\keV$). This requires a larger increase in
entropy, as the ambient entropy level at $0.15\rf$ in a $6\keV$ system
is $\sim300\keV\cc$ \citep{pra10}. This is achievable in principal
without additional feedback energy, via the amplification by merger
shocks of entropy input in cluster cores. This amplification is
particularly effective if the original entropy increase was driven by
lowering of the gas density rather than increasing its temperature
\citep[this effect is discussed in][in the context of the accretion of
material during cluster formation]{pon03}. With the entropy further
increased by merger shocks, the gas may then be able to settle to
radii well beyond the cluster core.  This model is similar to that
proposed by \citet{pra10}, but with an emphasis on the role of mergers
to amplify the entropy increase due to central energy input.

A challenge remains if we wish to explain all aspects of the observed
departures from strong self-similarity in our sample with this central
heating and merger shock mechanism: the loci of the $>6\keV$
unrelaxed/NCC clusters lie {\em above} the self-similar \LT\
relation. We thus speculate that while the combination of central
heating and merger shocks may act to raise gas entropy beyond
$0.15\rf$ in $\kT<6\keV$ systems, at higher masses the dominant effect
of mergers is to enhance \Lx\ and/or suppress kT.
We note that numerical simulations of clusters do not
unambiguously support this hypothesis. \citet{row04} found that
merging clusters tend to move along the \LT\ relation, though it was
observed that \Lx\ increases and kT decreases when the cores of two
systems merge. A similar effect was also present in the merger
simulations of \citet{poo07}, in which massive $\ga6\keV$ clusters
were shown to have enhanced \Lx\ relative to kT when the cores are at
their closest initial approach, before settling down onto the \LT\
relation over a period of $\approx4-5\Gyr$. This could explain the
observed position of the unrelaxed clusters above the self-similar
relation at the high-kT end, particularly combined with a selection
bias to preferentially detect and observe clusters that are at the
more luminous stages of their mergers (although we note that at a
close core passage, such clusters may not be identified as unrelaxed).

By contrast, in the cosmological simulations of \citet{har08}, mergers
were found to push clusters along, but slightly below the mean \LT\
relation, leading to a curve towards a flatter \LT\ relation for
disturbed clusters at the high kT end. This appears to contrast with
our results, but the comparison is not straightforward, as the
simulations are necessarily dominated by lower mass (kT$<3\keV$)
objects due to the limited volume simulated, and the classification of
disturbed clusters is based on time since the last major merger,
rather than a morphological measurement. For the same reasons, the
results of these different computational studies are not necessarily
at odds.

\subsection{Heating or \Lx\ suppression?}
Given the loci of lower mass systems below the self-similar \LT\
relation, it is interesting to consider whether those clusters occupy
that position due to their temperatures being enhanced relative to
their virial temperature, or their luminosities being suppressed due
to restructuring of the ICM. In some sense, this is just a question of
timescales, as heating of a parcel of gas will eventually \citep[on a
timescale of order the sound-crossing time: a few $10^{8}$
years;][]{mat11} have risen, expanding and cooling, to its new adiabat,
resulting in a restructuring of the ICM. Most theoretical work
focuses on this longer term result of the restructured ICM, but
some observational studies have argued that for clusters hosting
radio-loud AGN, the steepening of the \LT\ relation is due to
temperature enhancement of the ICM \citep{cro05,mag07}.

In the absence of a coherent set of radio data for these clusters, we
can appeal to their gas density profiles to investigate whether the
steepening of the \LT\ relation is due to enhanced kT or suppressed
\Lx\ of the ICM. Fig. \ref{f.cro08prof} shows that the ICM structure
of the lower mass CC clusters is different from the self-similar
higher mass CC profiles. The ICM density in the lower mass CC clusters
is lower than or similar to the self-similar profiles, suggesting that
the luminosity is suppressed in these systems by the removal of gas
from the central regions out to R500 and beyond.

For {\em unrelaxed} clusters, Fig. \ref{f.rho} shows that their ICM
structure is flatter than that of the self-similar relaxed
clusters. Indeed, the gas density is higher for unrelaxed clusters at
scaled radii $\gta0.4\rf$. A similar effect is seen in the scaled gas
density profiles of the {\em Planck}-selected sample of largely
unrelaxed clusters, when compared with the mean profile of the full
REXCESS sample \citep{pla11}. Indeed, the same excess of gas
density at large radii in NCC clusters is also present (albeit more
weakly) in the REXCESS profiles of \citep{cro08} if they are separated
into CC and NCC subsets. Given that the gas density is higher than the
self-similar profiles at large scaled radii, it is not immediately
clear if the net effect would be a suppression or increase of
\Lx. This can be crudely estimated by defining a ``psuedo-\Lx'' as the
integral of the square of the scaled mean density profile over the
spherical shell from $[0.15-1]\rf$. This calculation shows that the
unrelaxed clusters have a mean luminosity $\sim15\%$ lower than the
relaxed clusters, so that the net effect of the observed ICM
structural differences is suppression of \Lx.

\begin{figure}
\begin{center}
\scalebox{0.32}{\includegraphics*[angle=270]{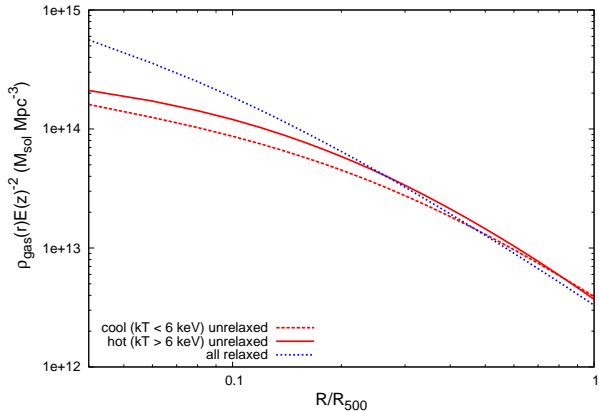}} \\
\caption[]{\label{f.mean_rho_kt} Average scaled gas density profiles of the
  relaxed and unrelaxed subsamples, with unrelaxed clusters subdivided into
  hot ($\kT>6\keV$) and cool ($\kT<6\keV$) subsets.}
\end{center}
\end{figure}

This picture changes slightly if the unrelaxed clusters are split into
hot ($\kT>6\keV$) and cool ($\kT<6\keV$) subsets. The mean profiles of
these subsets are shown in Fig. \ref{f.mean_rho_kt}. The temperature
dependence of the unrelaxed profiles is apparent, with cooler systems
having a flatter mean profile, which crosses the mean relaxed profile
at a larger scaled radius than the hotter unrelaxed systems. This is
consistent with the picture of core feedback resulting in the removal
of gas to larger radii in lower mass systems. The effect on the system
luminosity is also different. For the hotter unrelaxed clusters, the
psuedo-\Lx\ is $\sim5\%$ higher than the relaxed clusters, while for
the cooler unrelaxed clusters it is $\sim20\%$ lower. This is
qualitatively consistent with the crossover of the relaxed/CC and
unrelaxed/NCC \LT\ relations at $6\keV$. For a quantitative
comparison, the ratio of the luminosity predicted by the RCC
and NRCC relations (Fig. \ref{f.ltcc3}) was computed for the
temperature of each unrelaxed cluster. For the $\kT<6\keV$ clusters,
the mean ratio was $0.74$, indicating that the $20\%$ reduction in
\Lx\ suggested by the gas density profiles is sufficient to explain
the steeper unrelaxed/NCC relation, without enhancement of the ICM
temperature. For the $\kT>6\keV$ unrelaxed clusters, the mean ratio
was 1.30, significantly larger than the $5\%$ increase in \Lx\
predicted by the mean density profiles.

\subsection{Weak self-similarity}
Our investigation of the evolution of the \LT\ relation suggests at
first glance, a set of clusters at intermediate redshifts
$(0.2<z<0.6)$ with luminosities significantly higher than predicted by
self-similar evolution of the low-redshift relation (Figure
\ref{f.dl}). This is similar to the results of \citet{bra07}, who
argued in favour of non self-similar evolution. Our simple simulations
of a self-similar population of clusters subjected to selection
functions that plausibly approximate those of our heterogeneously
selected sample indicated that the deviations from self-similarity can
reasonably be explained by selection biases. As demonstrated by the
simulations, the combination of intrinsic scatter $\sigma_{L|M}$, and
the slope of the mass function at the mass corresponding to the survey
flux limit give differing amounts of bias for different flux limit and
redshift combinations. We thus find no evidence to reject the
self-similar description of evolution in the \LT\ relation. This
conclusion is in agreement with other analyses based on statistically
complete samples where full treatment of selection effects is possible
(\citet[][29 clusters at $z\lta1$]{pac07} \citet[][238 clusters at
$z\lta0.5$]{man10a}) which found no strong evidence for departures
from self-similar evolution in the \LT\ or \LM\ relation
\citep[although we note that in a similar analysis,][found an
indication of evolution slightly weaker than the self-similar
prediction for the \LM\ relation]{vik09}.

\begin{figure}
\begin{center}
\scalebox{0.6}{\includegraphics*[angle=0]{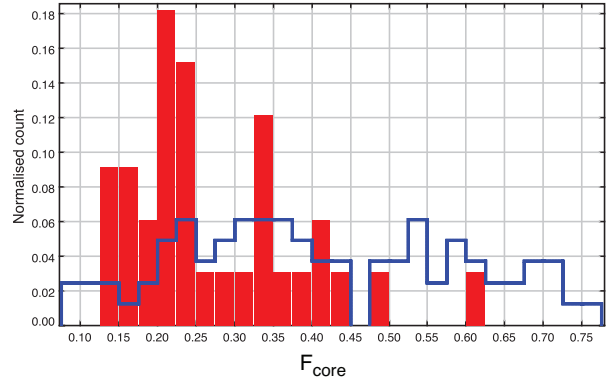}}
\caption[]{\label{f.fcore_histo} Histogram of \Fcore\ values for
  clusters at $z<0.5$ (lines) and $z\ge0.5$ (filled bars). Each
  histogram is normalised to unity.}
\end{center}
\end{figure}

Our simple simulations assumed $\sigma_{L|M}$ was constant with
redshift, but in fact this may not be the case. \citet{mau07b} showed
that $\sigma_{L|M}$ was significantly lower for the clusters in the
current sample at $z>0.5$ than those at $z<0.5$. This is consistent
with evidence that the fraction of clusters with cool cores (the
dominant contributor to $\sigma_{L|M}$) decreases strongly above
$z\approx0.5$ \citep{vik06c}. Evolution in $\sigma_{L|M}$ could
introduce further bias in measurements of the \LT\ relation
evolution. This evolution of the CC clusters was investigated in our
sample, and Figure \ref{f.fcore_histo} shows normalised histograms of
\Fcore\ for $z\ge0.5$ and $z<0.5$ clusters.

Fig. \ref{f.fcore_histo} shows a clear absence of strong CC
($\Fcore>0.5$) clusters at $z>0.5$ in the current sample. A similar
trend is seen in the simulations of \citet{kay07}, who found a
corresponding reduction in $\sigma_{L|T}$ with redshift. Our results
agree with the \citet{vik06c} findings, and we also found similar
histograms when using the cuspiness CC proxy. We also showed
previously that the fraction of relaxed clusters decreased
significantly at $z>0.5$ in the current sample
\citep{mau07b}. However, while these results provide strong evidence
for evolution in the CC population in our sample, we remind the reader
once more of the heterogeneous nature of our sample and urge caution
in extrapolating these results to the general cluster population
\citep[see e.g.][for discussions of cool core properties in
representative samples]{san09,mit11}. Indeed, even in statistically
complete samples, the fraction of CC clusters at high redshift is
subject of some debate. \citet{san10} recently showed that the cool
core fractions in three high-redshift \ROSAT-derived surveys differed
significantly, most likely due to their different cluster detection
algorithms (high-z cool core clusters could be classed as point
sources and excluded from samples). In two of the surveys considered,
the distribution of CC strengths (measured with a method similar to
our \Fcore\ proxy) was similar at low and high ($z>0.6$)
redshift. This would imply weaker (or zero) evolution in
$\sigma_{L|M}$ than found in our non-representative sample. The
uncertainty on the evolution in $\sigma_{L|M}$ further complicates
attempts to fully include selection bias corrections in studies of the
evolution in the \LT\ relation, although note that \citet{vik09}
showed that estimates of $\sigma_{L|M}$ from flux-limited samples are
unbiased, making bias-corrected studies feasible.

The redshift evolution of the relaxed/CC fraction in our sample means
that the relaxed/CC and unrelaxed/NCC subsets cover different redshift
ranges with the relaxed/CC subsets essentially limited to $z<0.5$. In
order to test whether this difference contributed to the shallower
\LT\ slopes found for the relaxed/CC clusters, the \LT\ relations of
the unrelaxed/NCC clusters were fit for $z<0.5$ and $z\ge0.5$ clusters
separately. There slopes of both the high and low redshift
unrelaxed/NCC clusters agreed very well with the fit to the full
unrelaxed/NCC subsample, and there was no evidence for a change of
slope with redshift.

\section{Conclusions}
We have used a sample of 114 clusters observed with \Chandra\ ACIS-I
across a wide baseline in temperature ($2<\kT<16\keV$) and redshift
($0.1<z<1.3$) to study the self similarity of the cluster population
in terms of its mass scaling and redshift evolution. Our main
conclusions on the mass scaling of the clusters are as follows:

\begin{itemize}
\item The ICM of massive ($\kT>3.5\keV$), relaxed/CC galaxy clusters is
  self-similar outside of the central $0.15\rf$. In this regime,
  strong self-similarity is obeyed, manifested in an \LT\ relation
  with a slope of $\approx2$, and ICM density profiles with low
  dispersion and no temperature dependence.
\item By comparing our data with measurements of the REXCESS sample,
  which extends to lower masses, we find that the self similarity of
  the relaxed/CC clusters breaks below $\sim3.5\keV$, manifested by a
  steepening \LT\ relation and flatter density profiles. This implies
  that the impact of central heating extends beyond the core in these
  lower mass systems.
\item Unrelaxed/NCC clusters are not self-similar;
  their \LT\ relation has a steeper than self-similar slope, and their
  ICM density profiles are temperature-dependent out to
  $\approx0.7\rf$.
\item The steeper unrelaxed/NCC core-excised \LT\ relation crosses the
  self-similar relaxed/CC relation at around $6\keV$. Below this
  temperature unrelaxed/NCC clusters appear to be less luminous than their
  relaxed/CC counterparts, while above this temperature they appear to
  be cooler and/or more luminous.
\item These results are consistent with similarity breaking in
  clusters being due to central feedback, the effects of which extend
  beyond the central $0.15\rf$ in low mass ($\lta3.5\keV$) systems. We
  suggest that merger shocks act to amplify the entropy increase from
  this feedback, allowing its effect to be felt beyond the core in more
  massive ($\lta6\keV$) unrelaxed systems. In the most massive
  systems, any heating is limited to the core regions and the dominant
  effect of mergers is to raise the luminosity of the ICM or lower its
  temperature relative to the self-similar \LT\ relation.
\end{itemize}

Our investigation of the evolution of the \LT\ relation gave the
following main results:
\begin{itemize}
\item The evolution of the \LT\ relation in our sample is inconsistent
  at face value with the self-similar model, but these differences are
  plausibly explained by a reasonable model of the varying selection
  biases due to the different selection functions in place across our
  heterogeneous sample.
\item Our bias modelling assumed the intrinsic scatter in the \LM\
  relation ($\sigma_{L|M}$) is redshift independent, however, in our
  sample the fraction of cool core clusters is much lower at $z>0.5$,
  consistent with the smaller $\sigma_{L|M}$ we found at $z>0.5$ for
  the same clusters in \citet{mau07b}. Such variation in
  $\sigma_{L|M}$ could further bias evolution measurements, but should
  be applied to the general population with caution due to the
  non-representative nature of our sample.
\item We would thus argue that for the core-excised \LT\ relation,
  self-similar evolution is obeyed to at least a first order
  approximation. This suggests that the balance of heating, cooling
  and mergers has remained roughly constant since $z\sim1$. Stronger
  constraints on non self-similar behaviour require statistically
  complete samples at all redshifts, with a full treatment of the
  selection function, including the sensitivity of the cluster
  detection algorithm to cool core clusters at high redshift.
\end{itemize}

Current X-ray surveys such as XXL \citep{pie10},
XCS \citep{llo10}, and XDCP \citep{fas08} are beginning
to produce suitable samples to study cluster evolution on a stronger
statistical foundation, extending the work of \citet{man10a} and
\citet{vik09} to $z>1$. Further insight will be gained by studies
of samples selected independently of the X-ray emission such as
optical, weak lensing, and to some extent the Sunyaev-Zel'dovich
effect (SZE; although the X-ray properties of SZE selected clusters
may retain some bias due to the dependence of both on the physical
properties of the ICM).

\section{Acknowledgements}
We thank Trevor Ponman, Alastair Sanderson and Gabriel Pratt for
useful discussions of this work. The financial support for SWR was
partially provided for by the Chandra X-ray Center through NASA
contract NAS8-03060, and the Smithsonian Institution.

\bibliographystyle{mn2e}
\bibliography{clusters}

\end{document}